\documentclass[british,english,aps]{revtex4}
\usepackage[T1]{fontenc}
\usepackage[latin9]{inputenc}
\usepackage{float}
\usepackage{amsmath}
\usepackage{amsthm}
\usepackage{graphicx}
\usepackage{esint}

\makeatletter
\numberwithin{equation}{section}
\numberwithin{figure}{section}

\usepackage{bbold}
\usepackage{braket}
\usepackage{color}
\usepackage{setspace}
\usepackage{mathrsfs}

\def\<{\langle}
\def\>{\rangle}

\def\beq{\begin{equation}}
\def\eeq{\end{equation}}
\def\barray{\begin{eqnarray}}
\def\earray{\end{eqnarray}}

\let\C\noexists

\newcommand{\C}{\mathbb{C}}
\newcommand{\Tr}{\text{Tr}}

\newcommand{\Rn}{\mathcal{R}_n}
\newcommand{\Ru}{\mathcal{R}_1}

\newcommand{\e}{\textit{e}}

\let\textquotedbl="
\renewcommand\[{\begin{equation}}
\renewcommand\]{\end{equation}} 

\@ifundefined{showcaptionsetup}{}{%
 \PassOptionsToPackage{caption=false}{subfig}}
\usepackage{subfig}
\makeatother

\usepackage{babel}
\begin{document}

\title{Unusual Corrections to the scaling of the entanglement entropy of the excited
states in conformal field theory}

\author{Lorenzo Cevolani }

\affiliation{Laboratoire Charles Fabry, Institut d'Optique, CNRS, Univ. Paris-Saclay, 2 avenue Augustin Fresnel, F-91127 Palaiseau cedex, France}
\begin{abstract}
In this paper we study the scaling of the correction of the Rényi entropy
of the excited states in systems described, in the continuum limit, by a conformal field theory (CFT). These corrections
scale as $L^{-\frac{2\Delta}{n}}$, where $L$ is the system size
and $\Delta$ is the scaling dimension of a relevant bulk operator
located around the singularities of the Riemann surface $\Rn$. Their name is due to their explicit dependence on the Riemann surface $\Rn$.
Their presence has been detected in several works on the entanglement entropy in finite size systems, both in the ground and the excited states. Here, we present a general study of these corrections based on the perturbation expansion on $\Rn$. Some of the terms in this expansion are divergent and they will be cured with addition cut-offs. These cut-offs will determine how these corrections scale with the system size $L$. Exact
numerical computations of the Rényi entropy of the excited states of the
XX model are provided and they confirm our theoretical prediction on
the scaling of corrections. They allow also a comparison with the other
works present in the literature finding that the corrections, for the excited states, have the exact same form of the ones of the ground state case multiplied by a model dependend function of $n$ and $l/L$.
\end{abstract}
\maketitle

\section{Introduction}

From the birth of the quantum theory, the entanglement has been one of
the more fascinating features of the new description of the microscopic
world. The presence of a spooky ``action at distance'' without a
classical counterpart, Ref.~\cite{scrod1,scrod2,gatto}, has raised endless
discussions between the fathers of the newborn theory and it has been
defined as ``the'' nature of the quantum mechanics itself, Ref.~\cite{EPR}.
After more than a century, entanglement is still a key quantity to
understand many phenomena from condensed matter theory to cosmology,
Ref.~\cite{RevModPhys81865,Black1,Black2,Black3,algoritmodishor}. The Rényi entropy
$S_{A}^{(n)}=\left(\frac{1}{1-n}\right)\ln\Tr\rho_{A}^{n}$ has been
found to be one of the best ways to quantify the amount of entanglement, see Ref.~\cite{entanglementmagnetic,RevModPhys80517},
between a system $S$ in a pure state and a subsystem $A$ described by the reduced density matrix $\rho_A$. The limit
$n\rightarrow1^{+}$ gives the Von Neumann entropy $S_{A}=-\Tr\rho_{A}\log\rho_{A}$.
In general, the computation of $S_{A}^{(n)}$ or $S_{A}$ is an impossible
task: the complexity of the density matrix of an interacting system does
not allow any analytic computations. In the special case of a one dimensional system $S$, whose underlying continuum theory is a CFT, it is possible to compute the Rényi entropy for a generic index $n$. In the case where $S$ is in its ground state the Rényi entropy reads

\[
S_{CFT}^{(n)}=\frac{c}{6}\left(1+\frac{1}{n}\right)\ln\left[\frac{L}{\pi a}\sin\left(\frac{\pi l}{L}\right)\right]\stackrel{L\rightarrow \infty}{\sim}
\frac{c}{6}\left(1+\frac{1}{n}\right)\ln\left[\frac{l}{a}\right]
\]
where $L$ ($l$) is the length of $S$ ($A$), $a$ is an infrared
cut-off and $c$ is the central charge of the CFT, Ref.~\cite{DiFrancesco}. This expression has been found in several works, see Ref.~\cite{Holzey,Latorre2,nontecnica,PasqualeCFT,PasqualeEEN,351882935}, its universal scaling represents one of the most beautiful
consequences of the concept of universality in physics.
The previous expression is obtained using a multi-copy system defined
by $n$ copies of the initial one. These copies are connected one
to the other creating a new system that can be seen as defined on a complex Riemann surface $\Rn$, see Ref.~\cite{CardyDoyon}.
The same geometrical approach works also if $S$ is no longer in its ground state but in one of its excited states, Ref.~\cite{Berganza,BerganzaPRL}. The difference between the entropy of excited
state $S_{\Upsilon}^{(n)}$ and the one of the ground state $S_{gs}^{(n)}$
seen before can be computed again using the multi-copy model on the surface $\Rn$
\begin{equation}
\e^{(1-n)\left(S_{\Upsilon}^{n}-S_{gs}^{n}\right)}=\lim_{w\rightarrow-\imath\infty}\dfrac{\langle\prod_{j=0}^{n-1}\Upsilon_{j}(w)\Upsilon_{j}^{\dagger}(-w)\rangle_{\mathcal{R}_{n}}}{\left[\langle\Upsilon_{0}(w)\Upsilon_{0}^{\dagger}(-w)\rangle_{\mathcal{R}_{1}}\right]^{n}}\label{eq:in}
\end{equation}
where the notation $\langle\rangle_{\Rn}$ means an expectation value computed over the $n$-sheeted
Riemann surface.\\
The previous expressions have been proved to be true in several systems,
both analytically and numerically, see Ref.~\cite{Latorre2,Latorre,351882944,351882945,351882956,351882948,PhysRevA74022329,log1,igloi,log2,log3,Peschel}.
Furthermore, some fast oscillating geometry dependent corrections to the continuum
CFT results have been found and studied in different systems 
Ref.~\cite{steelisa,massiveunusual,PhysRevLett104095701,EsslerCalabrese,unusual2,unusual3,Taddia,Dalmonte1,Dalmonte2,Maurizio,Xavier,unusual4}.
In particular, the scaling of these corrections has been found to be $L^{-\frac{2\Delta}{n}}$ were $\Delta$
is the scaling dimension of a relevant bulk operator. This behavior is different
from the standard renormalization group (RG) flows which is $L^{-(\Delta-2)}$.
The understanding of these corrections is crucial to compare the continuum
limit and the numerical data from finite size systems. In Ref.~\cite{UnusualCorrections}
it was demonstrated that, for the ground state, these geometry-dependent
corrections are due to the presence of a relevant bulk operator $\Delta>2$ localized around the conical singularities
that describe the surface $\Rn$. In this paper we will extend this
result to the case of the excited states, where some oscillating corrections
to the CFT prediction have been found but not yet studied from the theoretical
point of view.\\
This work is organized as follows: in the first section we will study
how Eq.~(\ref{eq:in}) is affected by a perturbation to the conformal
invariant action. The entanglement entropy can then be evaluated as a
power series in the standard perturbation theory on $\Rn$. We will
look for divergences in the terms of this expansion and we will cure
them with additional cut-offs. This procedure will lead to the scaling
behavior of the corrections to the CFT predictions.\\
In the second part we will present the numerical results. We will study
the entanglement entropy of the particle-hole excitations of the XX
model, which corresponds to the operator $\imath\partial\phi$ in
the continuum limit of this model. We will confirm that the difference between the
entropy of this state and the one predicted by the CFT scales as $L^{-\frac{2}{n}}$.
We will then compare the corrections in the excited state case with
the one found in Ref.~\cite{correzzioni,correzzioni2} for the ground state finding that they have
the same dependence on the filling. We will conclude that the corrections in the excited states can be written as the ground state ones multiplied by a model dependent function $R_n(l/L)$ which was an unknown result before our analysis.

\section{Unusual corrections}

\subsection{Perturbation of the function $F_{\Upsilon}^{(n)}$}

We will study the unusual corrections to the scaling of the Rényi
entropy of the excited states produced by a primary field $\Upsilon(w)$ defined on $\Rn$.
The replica trick approach found for the ground state in Ref.~\cite{PasqualeCFT} can be extended to the excited states taking as "in" state
\begin{equation}
\ket{\Upsilon}=\lim_{(z;\bar{z})\rightarrow0}\Upsilon(z;\bar{z})\ket{0},
\end{equation}
where $\ket{0}$ is the standard vacuum state, and as "out" state
\begin{equation}
\bra{\Upsilon}=\lim_{(z;\bar{z})\rightarrow0}z^{-2h}\bar{z}^{-2\bar{h}}\bra{0}\Upsilon\left(\frac{1}{z};\frac{1}{\bar{z}}\right).\end{equation}
In the previous expressions we used radial quantization, see Ref.~\cite{DiFrancesco}, to define the infinite past and infinite future.\\
The sub-system $A$ is a segment of length $l<L$ inside the system $S$ and the Riemann surface $\Rn$, defined using the replica trick approach, is mapped onto $\C$ for every value of $n$ thanks to the conformal transformation
\begin{equation}
z=\left[\frac{\sin\left(\frac{\pi w}{L}\right)}{\sin\left(\frac{\pi(w-l)}{L}\right)}\right]^{\frac{1}{n}}\label{eq:confmap}
\end{equation}
Calling $\rho_{\Upsilon}^{n}$ the density matrix of the excited state and $\rho_{I}^{n}$ the one of the ground state, following 
Ref.~\cite{Berganza,BerganzaPRL}, the Rényi entropy can be written as
\begin{equation}
F_{\Upsilon}^{n}(x)\equiv\dfrac{Tr\rho_{\Upsilon}^{n}}{Tr\rho_{I}^{n}}=\exp\left((1-n)\left(S_{\Upsilon}^{n}-S_{gs}^{n}\right)\right)=\lim_{w\rightarrow-\imath\infty}\dfrac{\langle\prod_{j=0}^{n-1}\Upsilon_{j}(w)\Upsilon_{j}^{\dagger}(-w)\rangle_{\mathcal{R}_{n}}}{\left[\langle\Upsilon_{0}(w)\Upsilon_{0}^{\dagger}(-w)\rangle_{\mathcal{R}_{1}}\right]^{n}},\label{funzioneeffe}
\end{equation}
the limit $w\rightarrow -\imath \infty$ fixes the the operators $\Upsilon^\dagger(w)$ and $\Upsilon(w)$ to the infinite future and the infinite past. 
The correlation functions are evaluated using the path integral formulation of quantum mechanics and the conformal invariant action $S_0$.\\ 
Following now Ref.~\cite{UnusualCorrections} we want to study the response of the previous expression to a relevant perturbation $\Phi(w)$ with scaling dimensions $\Delta>2$ localized around the conical singularities of the Riemann surface. These conical singularities are identified on $\Rn$ by the points $w=0$ and $w=l$ and they can be mapped into the complex plane using the mapping \ref{eq:confmap}. The introduction of this perturbation operator changes the action to
\begin{equation}
S=S_{0}+\lambda\int_{\Rn}d^{2}w\Phi(w),\label{azioneperturbatissima}
\end{equation}
Consequently, \ref{funzioneeffe} can be formally written as
\begin{equation}
\tilde{F}_{\Upsilon}^{n}(x)=\lim_{w\rightarrow-\imath\infty}\dfrac{\langle\prod_{j=0}^{n-1}\Upsilon_{j}(w)\Upsilon_{j}^{\dagger}(-w)\e^{-\lambda\int_{\Rn}d^{2}w\,\Phi(w)}\rangle_{\mathcal{R}_{n}}}{\left[\langle\Upsilon_{0}(w)\Upsilon_{0}^{\dagger}(-w)\e^{-\lambda\int_{\Ru}d^{2}w\Phi(w)}\rangle_{\mathcal{R}_{1}}\right]^{n}},
\end{equation}
where the correlation functions are here evaluated using the CFT invariant action $S_{0}$.\\
The previous expression can be evaluated as power series of the coupling constant $\lambda \ll 1$ which can be written as $g/\epsilon^{2-\Delta}$ where $g$ is the dimensionless coupling constant, see Ref.~\cite{DiFrancesco}. Taking the logarithm of this expansion we will restrict ourself to the evaluation of just connected Feynman diagrams
\begin{eqnarray}
\ln\left(\tilde{F}_{\Upsilon}^{n}(x)\right) & = & \ln\left(F_{\Upsilon}^{n}(x)\right)+\label{perturbativeexpansion}\\
 &  & +\sum_{k=1}^{\infty}\dfrac{(-\lambda)^{k}}{k!}\idotsint_{\Rn}\dfrac{\langle\prod_{j}\Upsilon_{j}(w)\Upsilon_{j}^{\dagger}(-w)\Phi(w_{1})\Phi(w_{2})\dots\Phi(w_{k})\rangle_{\mathcal{R}_{n}}}{\langle\prod_{j=0}^{n-1}\Upsilon_{j}(w)\Upsilon_{j}^{\dagger}(-w)\rangle_{\mathcal{R}_{n}}}d^{2}w_{1}\dots d^{2}w_{k}\nonumber \\
 &  & -n\left(\sum_{k=1}^{\infty}\dfrac{(-\lambda)^{k}}{k!}\idotsint_{\Ru}\dfrac{\langle\Upsilon_{0}(w)\Upsilon_{0}^{\dagger}(-w)\Phi(w_{1})\Phi(w_{2})\dots\Phi(w_{k})\rangle_{\mathcal{R}_{1}}}{\langle\Upsilon_{0}(w)\Upsilon_{0}^{\dagger}(-w)\rangle_{\mathcal{R}_{1}}}d^{2}w_{1}\dots d^{2}w_{k}\right).\nonumber 
\end{eqnarray}
Some terms in the previous expansion have divergences due to the field $\Phi(w)$, as we will see, and they have to be cured imposing a short distance cut-off $\epsilon$. This 
corresponds to assume that the continuum theory describes the correct behavior of the correlations just on distances larger than $\epsilon$ because for smaller ones the continuum theory does not hold. This approximation is rude but it has been demonstrated to be sufficient to 
find the correct scaling of the corrections in the ground state case, Ref.~\cite{UnusualCorrections}.\\ 
We will study separately the two integrals defining the order $\lambda$ term in Eq.~\ref{perturbativeexpansion} 
\begin{equation}
\lambda\int_{\Rn}\dfrac{\langle\prod_{j=0}^{n-1}\Upsilon_{j}(w)\Upsilon_{j}^{\dagger}(-w)\Phi(w_{1})\rangle_{\mathcal{R}_{n}}}{\langle\prod_{j=0}^{n-1}\Upsilon_{j}(w)\Upsilon_{j}^{\dagger}(-w)\rangle_{\mathcal{R}_{n}}}d^{2}w_{1}-\lambda\int_{\Ru}\dfrac{\langle\Upsilon_{0}(w)\Upsilon_{0}^{\dagger}(-w)\Phi(w_{1})\rangle_{\Ru}}{\langle\Upsilon_{0}(w)\Upsilon_{0}^{\dagger}(-w)\rangle_{\Ru}}d^{2}w_{1},
\end{equation}
 and the same for the order $\lambda^2$
\begin{equation}
\lambda^{2}\int_{\Rn}\dfrac{\langle\prod_{j=0}^{n-1}\Upsilon_{j}(w)\Upsilon_{j}^{\dagger}(-w)\Phi(w_{1})\Phi(w_{2})\rangle_{\Rn}}{\langle\prod_{j=0}^{n-1}\Upsilon_{j}(w)\Upsilon_{j}^{\dagger}(-w)\rangle_{\Rn}}-\lambda^{2}\int_{\Ru}\dfrac{\langle\Upsilon_{0}(w)\Upsilon_{0}^{\dagger}(-w)\Phi(w_{1})\Phi(w_{2})\rangle_{\Ru}}{\langle\Upsilon_{0}(w)\Upsilon_{0}^{\dagger}(-w)\rangle_{\Ru}}.
\end{equation}

\subsection{The $\lambda$ term on the surface $\Ru$}
We start from the order $\lambda$ term defined on the surface $\Ru$ which reads
\begin{equation}
\int_{\Ru} d^2w_1\lim_{w\rightarrow-\imath\infty}\dfrac{\langle\Upsilon_{0}(w)\Upsilon_{0}^{\dagger}(-w)\Phi(w_{1})\rangle_{\Ru}}{\langle\Upsilon_{0}
(w)\Upsilon_ { 0 } ^ { \dagger}(-w)\rangle_{\Ru}}.\label{3p}
\end{equation}
Without loss of generality, we will assume that $\Phi$ is spinless, $\bar{h}_\Phi=h_\Phi$, and $\Upsilon$ has just an holomorphic part, $h_\Upsilon=h$ 
and $\bar{h}_\Upsilon=0$. Using the conformal map \ref{eq:confmap}, the integrand can be mapped from $\Ru$ to $\C$ using the standard transformation rule for conformal fields, see Ref.~\cite{DiFrancesco},
\begin{equation}
\lim_{w\rightarrow-\imath\infty}\dfrac{\langle\Upsilon_{0}(w)\Upsilon_{0}^{\dagger}(-w)\Phi(w_{1})\rangle_{\Ru}}{\langle\Upsilon_{0}(w)\Upsilon_{0}^{\dagger}(-w)\rangle_{\Ru}}=\left|\left(\dfrac{L\sin(\pi x)}{\pi}\right)\dfrac{1}{\left(e^{\imath\pi x}-z\right)\left(e^{-\imath\pi x}-z\right)}\right|^{-2h_{\Phi}}\dfrac{\langle\Upsilon(z_{out})\Upsilon^{\dagger}(z_{in})\Phi(z_{1})\rangle_{\C}}{\langle\Upsilon(z_{out})\Upsilon^{\dagger}(z_{in})\rangle_{\C}},\label{aa2}
\end{equation}
the points $z_{in}$ and $z_{out}$ are the "in" and "out" states mapped on the complex plane from the Riemann surface. The previous expression can be computed explicitly thanks to the conformal invariance on $\C$
\begin{equation}
\dfrac{\langle\Upsilon_{0}(z^{\prime})\Upsilon_{0}^{\dagger}(z)\Phi(z_{1})\rangle_{\C}}{\langle\Upsilon_{0}(z^{\prime})\Upsilon_{0}^{\dagger}(z)\rangle_{\C}}=\left(\dfrac{\left|2\sin(\pi x)\right|}{\left|e^{\imath\pi x}-z_{1}\right|\left|e^{-\imath\pi x}-z_{1}\right|}\right)^{2h_{\Phi}}.
\end{equation}
Even if this expression depends on the variable $z_1$ on the complex plane, its dependence is canceled out by the transformation factor from $\Rn$ to $\C$. The integrand of \ref{3p} is then a constant and it has no divergences around the conical singularities
\[
\lim_{w\rightarrow-\imath\infty}\dfrac{\langle\Upsilon_{0}(w)\Upsilon_{0}^{\dagger}(-w)\Phi(w_{1})\rangle_{\Ru}}{\langle\Upsilon_{0}(w)\Upsilon_{0}^{
\dagger}(-w)\rangle_{\Ru}}=\left(\dfrac{2\pi}{L}\right)^{2h_{\Phi}}.
\]
No further cut-off of order $\epsilon$ is needed and the scaling of this term is given by the coupling constant $\lambda=g/\epsilon^{2-\Delta}$.

\subsection{The $\lambda^{2}$ term on the surface $\Ru$}

On the same Riemann surface $\Ru$ we can study the next order $\lambda^{2}$ of \ref{perturbativeexpansion} which reads
\begin{equation}
\iint_{\Ru} 
d w_1 
d w_2\lim_{w\rightarrow-\imath\infty}\dfrac{\langle\Upsilon(w^{\prime})\Upsilon^{\dagger}(w)\Phi(w_{1})\Phi(w_{2})\rangle_{\Ru}}{\langle\Upsilon(w^{ 
\prime })\Upsilon^{\dagger}(w)\rangle_{\Ru}},\label{lolo}
\end{equation}
as usual the integrand can be mapped from this Riemann surface to $\C$ thanks to conformal symmetry
\begin{equation}
\dfrac{\langle\Upsilon(w^{\prime})\Upsilon^{\dagger}(w)\Phi(w_{1})\Phi(w_{2})\rangle_{\Ru}}{\langle\Upsilon(w^{\prime})\Upsilon^{\dagger}(w)\rangle_{\Ru}}=\left|\dfrac{dw_{1}}{dz_{1}}\right|^{-2h_{\Phi}}\left|\dfrac{dw_{2}}{dz_{2}}\right|^{-2h_{\Phi}}\dfrac{\langle\Upsilon(z_{0}^{\prime})\Upsilon^{\dagger}(z_{0})\Phi(z_{1})\Phi(z_{2})\rangle_{\C}}{\langle\Upsilon(z_{0}^{\prime})\Upsilon^{\dagger}(z_{0})\rangle_{\C}},
\end{equation}
but in this case, the conformal invariance is not sufficient to fix completely the four point function on the complex plane 
\begin{equation}
\dfrac{\langle\Upsilon_{0}(w^{\prime})\Upsilon_{0}^{\dagger}(w)\Phi(w_{1})\Phi(w_{2})\rangle_{\Ru}}{\langle\Upsilon_{0}(w^{\prime})\Upsilon_{0}^{\dagger}(w)\rangle_{\Ru}}=\left|\dfrac{dw_{1}}{dz_{1}}\right|^{-2h_{\Phi}}\left|\dfrac{dw_{2}}{dz_{2}}\right|^{-2h_{\Phi}}|z_{1}-z_{2}|^{-4h_{\Phi}}F\left(\eta;\bar{\eta}\right).\label{eq:4pti/2pti}
\end{equation}
where $F(\eta;\bar{\eta})$ is an invariant continuous scalar function of the anharmonic ratio
\begin{equation}
\eta=\dfrac{(z_{0}^{\prime}-z_{1})(z_{0}-z_{2})}{(z_{0}^{\prime}-z_{2})(z_{0}-z_{1})}.
\end{equation}
Eq.~\ref{eq:4pti/2pti} exhibits the same dependence on the variables $z_{1}$ and $z_{2}$ as the correlator $\langle \Phi(w_1)\Phi(w_2) \rangle$ in the ground state case Ref.~\cite{UnusualCorrections}, but here it is multiplied by an invariant continuous function of $\eta$. In the limit where $z_1$ and $z_2$ approach together the conical singularities on $\C$, 
the function $F(\eta;\bar{\eta})$ tends to a constant thanks to its continuity, see Ref.~\cite{DiFrancesco}. Its contribution to Eq.~\ref{lolo} is an overall constant factor which can be ignored in the followings.\\
We can then study the integral \ref{eq:4pti/2pti} using the explicit expression found for the integrand and the fact that 
$F\left(\eta:\bar{\eta}\right)$ is a approximatively constant in this region
\begin{equation}
\frac{F\left(\eta;\bar{\eta}\right)}{2}\lambda^{2}\int_{\C}dz_{1}\int_{\C}dz_{2}\left|\dfrac{dw_{1}}{dz_{1}}\right|^{-2h_{\Phi}}\left|\dfrac{dw_{2}}{dz_{2}}\right|^{-2h_{\Phi}}|z_{1}-z_{2}|^{-4h_{\Phi}}.\label{eq:int1}
\end{equation}
The contribution from the region $z_1\sim 0$ and $z_2 \rightarrow \infty$ can be easily computed because the integrand factorizes and we can 
evaluate separately the two contributions from the two conical singularities. For the region $z_1\sim 0$ 
\begin{equation}
\int_{\C}d^{2}z_{1}\left|\dfrac{dw_{1}}{dz_{1}}\right|^{2-2h_{\Phi}}\approx\int_{\Ru}d^{2}w_{1}\left|\dfrac{\pi}{L\sin(\pi x)}\right|^{2h_{\Phi}}=2\pi\left|\dfrac{L\sin(\pi x)}{\pi}\right|^{-2h_{\Phi}}\epsilon^{2}.
\end{equation}
where the prefactor comes out from the linearization of \ref{eq:confmap} around the conical singularity $w = 0$. This integral is divergent and we need to impose a cut-off of order $\epsilon$. The same procedure has to be done for the integral on $z_{2}$ approximated around $w \approx l$
\begin{equation}
\int_{\C}d^{2}z_{2}\left|\dfrac{dw_{2}}{dz_{2}}\right|^{2}\left|\dfrac{dz_{2}}{dw_{2}}\right|^{2h_{\Phi}}\left|z_{2}\right|^{-4h_{\Phi}}=\left|\dfrac{L\sin(\pi x)}{\pi}\right|^{-2h_{\Phi}}\int_{\Ru}d^{2}w_{2}=2\pi\left|\dfrac{L\sin(\pi x)}{\pi}\right|^{-2h_{\Phi}}\epsilon^{2}.
\end{equation}
This expression needs to be further regulated imposing a cut-off of order $1/\epsilon$ on the upper bound. Taking into account the contribution coming from the integral and from the factor $\lambda^2$ we find the scaling of the term
\begin{equation}
\left(\dfrac{L\sin(\pi x)}{\pi}\right)^{-2\Delta},
\end{equation}
which is the scaling of the unusual corrections on the Riemann surface $\Ru$.

\subsection{The $\lambda^{2}$ term on the surface $\Rn$}

We want to continue our analysis with the term $\lambda^2$ defined on the Riemann surface $\Rn$ in \ref{perturbativeexpansion}
\begin{equation}
\dfrac{\lambda^{2}}{2}\int_{\Rn}d^{2}w_{1}\int_{\Rn}d^{2}w_{2}\lim_{w\rightarrow\imath\infty}\dfrac{\langle\prod_{j=0}^{n-1}\Upsilon_{j}(w)\Upsilon_{j}^{\dagger}(-w)\Phi(w_{1})\Phi(w_{2})\rangle_{\mathcal{R}_{n}}}{\langle\prod_{j=0}^{n-1}\Upsilon_{j}(w)\Upsilon_{j}^{\dagger}(-w)\rangle_{\mathcal{R}_{n}}}.\label{ouououou}
\end{equation}
The expectation value of an arbitrarily long string of different operators can be computed with a trick: we compute the correlator replacing all the operators with vertex operators with the same holomorphic and antiholomorphic dimensions as the original ones. Then we multiply this result by a scalar function of all the possible anharmonic rations. The simplification comes from the fact that the correlator of the product
of vertex operators can be computed thanks to the conformal symmetry
\begin{equation}
\langle\prod_{i=0}^{n-1}\mathcal{V}_{\alpha_{i}}(z_{i})\rangle_{\C}=\prod_{i<j}\left|z_{i}-z_{j}\right|^{4\alpha_{i}\alpha_{j}}
\end{equation}
if the neutrality condition $\sum_{i}\alpha_{i}=0$ is fulfilled.
The $\alpha$ parameter of the vertex operator is connected to the scaling dimension  as $\alpha^2=h=\bar{h}$. For all the vertex operators, we will tune this parameter to match the scaling dimensions of the original operators, and at the same time to satisfy the neutrality condition. For example, for the field $\Phi$ we have $h_{\Phi}=\bar{h}_{\Phi}$. Therefore, we will replace it with a vertex operator whose $\alpha$ is
\begin{equation}
\alpha_{\Phi}=\pm\sqrt{h_{\Phi}}.\label{alfatoh}
\end{equation}
For the $\Upsilon$ operator the situation is slightly different. This operator has only the holomorphic part $h_{\Upsilon}=h$ different from zero. We can consider a vertex operator with $\bar{\alpha}=0$ and $\alpha=\pm\sqrt{h}$,
which means simply that the correlator between these vertex operators
has a vanishing anti-holomorphic part. The signs of the $\alpha$ parameters are chosen to satisfy
the neutrality condition for the numerator and the denominator of \ref{ouououou} at the same time.
The fields $\Upsilon$ are replaced by vertex operators with parameter
\begin{equation}
(\alpha_{\Upsilon};\bar{\alpha}_{\Upsilon})=(\sqrt{h};0)\label{neutrality1}
\end{equation}
which satisfies the neutrality condition for the denominator alone in Eq.~\ref{ouououou}.
For the fields $\Phi(z_1)$ and $\Phi(z_2)$ we use vertex operators defined by $\alpha_{\Phi_{1}}$
and $\alpha_{\Phi_{2}}$ opposite to each others as 
\begin{eqnarray}
\left(\alpha_{\Phi_{1}};\bar{\alpha}_{\Phi_{1}}\right) = -\left(\alpha_{\Phi_{2}};\bar{\alpha}_{\Phi_{2}}\right)& = & \left(\sqrt{h_{\Phi}};-\sqrt{h_{\Phi}}\right).
\end{eqnarray}
This choice, together with the previous one, satisfies the neutrality condition for the numerator. We can now compute the two correlators involving the vertex operators
\begin{eqnarray}
 & \dfrac{\langle\prod_{i=0}^{n-1}\mathcal{V}_{\alpha_{i}}(z_{i}^{\prime})\mathcal{V}_{\alpha_{i}}^{\dagger}(z_{i})\mathcal{V}_{\alpha_{\Phi_{1}}}(z_{1})\mathcal{V}_{\alpha_{\Phi_{2}}}(z_{2})\rangle_{\C}}{\langle\prod_{i=0}^{n-1}\mathcal{V}_{\alpha_{i}}(z_{i}^{\prime})\mathcal{V}_{\alpha_{i}}^{\dagger}(z_{i})\rangle_{\C}}=\left[\prod_{i=0}^{n-1}\dfrac{(z_{i}^{\prime}-z_{1})(z_{i}^{\prime}-z_{2})}{(z_{i}-z_{1})(z_{i}-z_{2})}\right]^{2\sqrt{hh_{\Phi}}}\dfrac{1}{\left|z_{1}-z_{2}\right|^{4h_{\Phi}}}. 
\end{eqnarray}
The integrand of Eq.~\ref{ouououou} is then given by the product between the previous expression and an unknown invariant function of all the anharmonic ratios
\begin{align}
 & \dfrac{\langle\prod_{j=0}^{n-1}\Upsilon_{j}(w)\Upsilon_{j}^{\dagger}(-w)\Phi(w_{1})\Phi(w_{2})\rangle_{\mathcal{R}_{n}}}{\langle\prod_{j=0}^{n-1}\Upsilon_{j}(w)\Upsilon_{j}^{\dagger}(-w)\rangle_{\mathcal{R}_{n}}}=\label{eq:corrr}\\
= & \left|\dfrac{dw_{1}}{dz_{1}}\right|^{-2h_{\Phi}}\left|\dfrac{dw_{2}}{dz_{2}}\right|^{-2h_{\Phi}}F(\eta_{i})\left(\prod_{i=0}^{n-1}\dfrac{(z_{j}^{\prime}-z_{1})(z_{j}^{\prime}-z_{2})}{(z_{j}-z_{1})(z_{j}-z_{2})}\right)^{2\sqrt{hh_{\Phi}}}\dfrac{1}{\left|z_{1}-z_{2}\right|^{4h_{\Phi}}}.
\end{align}
where $\eta_{j}$ are the anharmonic ratios involving the variables $z_{1}$ and $z_{2}$
\begin{equation}
\eta_{j}=\dfrac{(z_{j}^{\prime}-z_{1})(z_{j}-z_{2})}{(z_{j}^{\prime}-z_{2})(z_{j}-z_{1})}.
\end{equation}
Expression \ref{eq:corrr} depends $z_{1}$ and $z_{2}$ exactly in the same way
as $\langle\Phi(z_{1})\Phi(z_{2})\rangle$ but multiplied by the invariant function
\begin{equation}
F(\eta_{i})\left(\prod_{i=0}^{n-1}\dfrac{(z_{j}^{\prime}-z_{1})(z_{j}^{\prime}-z_{2})}{(z_{j}-z_{1})(z_{j}-z_{2})}\right)^{2\sqrt{hh_{\Phi}}}.
\end{equation}
This expression becomes an overall constant in the limit where $z_{1}\rightarrow 0$
and $z_{2}\rightarrow \infty$
\begin{eqnarray}
 & \eta_{j} & =\dfrac{(z_{j}^{\prime}-z_{1})(z_{j}-z_{2})}{(z_{j}^{\prime}-z_{2})(z_{j}-z_{1})}\rightarrow\dfrac{z_{j}^{\prime}}{z_{j}},\\
 & F & (\eta_{i})\left(\prod_{i=0}^{n-1}\dfrac{(z_{j}^{\prime}-z_{1})(z_{j}^{\prime}-z_{2})}{(z_{j}-z_{1})(z_{j}-z_{2})}\right)^{2\sqrt{hh_{\Phi}}}\rightarrow F\left(\dfrac{z_{j}^{\prime}}{z_{j}}\right)\prod_{j=0}^{n-1}\left(\dfrac{z_{j}^{\prime}}{z_{j}}\right)^{2\sqrt{hh_{\Phi}}}.
\end{eqnarray}
This contributes to the integral as an overall constant and we can ignore it in the followings. The contribution to the integral \ref{ouououou} coming from the conical singularities, $w_1\sim 0$ and $w_2\sim l$ on the Riemann surface, can be computed as
\begin{align}
 & \int_{\C}d^{2}z_{1}\int_{\C}d^{2}z_{2}\left|\dfrac{dw_{1}}{dz_{1}}\right|^{2-2h_{\Phi}}\left|\dfrac{dw_{2}}{dz_{2}}\right|^{2-2h_{\Phi}}\dfrac{1}{\left|z_{1}-z_{2}\right|^{4h_{\Phi}}}=\nonumber \\
 & =\int_{\C}d^{2}z_{1}\left|\dfrac{dw_{1}}{dz_{1}}\right|^{2-2h_{\Phi}}\int_{\C}d^{2}z_{2}\left|\dfrac{dw_{2}}{dz_{2}}\right|^{2-2h_{\Phi}}\left|z_{2}\right|^{-4h_{\Phi}}=\nonumber \\
 & =\left|\frac{nL\sin(\pi x)}{\pi}\right|^{-\frac{4h_{\Phi}}{n}}\int_{\Rn}d^{2}w_{1}\left|w_{1}\right|^{(1-\frac{1}{n})2h_{\Phi}}\int_{\Rn}d^{2}w_{2}\left|w_{2}-l\right|^{(\frac{1}{n}-1)2h_{\Phi}}.
\end{align}
These integrals are both divergent for $n>n_{c}$ where $n_{c}=\Delta/\left(\Delta-2\right)$, which is the same result found in Ref.~\cite{UnusualCorrections} for the ground state. We impose then two cut-offs of order $\epsilon$ in the $\Rn$ space to cure these divergences
\begin{equation}
\left|\frac{nL\sin(\pi x)}{\pi}\right|^{-\frac{4h_{\Phi}}{n}}(2\pi)^{2}\epsilon^{4h_{\Phi}(\frac{1}{n}-1)},
\end{equation}
taking now into account the $\epsilon$ dependence of the coupling constant $\lambda^2$ given by the RG we get the final result:
\begin{equation}
\left(\dfrac{L\sin(\pi x)}{\pi}\right)^{-\frac{2\Delta}{n}}.
\end{equation}
We have then found that the geometry dependent corrections to the scaling of the entanglement entropy computed using the CFT are present also when the systems is in one of its excited states. The scaling of this corrections is the same one already found in the ground state and it is due to their purely geometrical nature.

\subsection{The $\lambda$ term on the surface $\Rn$}

We can now analyze the order $\lambda$ term defined on the surface $\Rn$ in \ref{perturbativeexpansion}. We will perform the calculation for a specific model: the bidimensional critical Ising model, this is because the method used in the previous section cannot be applied here. We choose $\Upsilon(w_{i})=\sigma(w_{i})$ as the excitation operator and $\Phi(w_{1})=\epsilon(w_{1})$ as the perturbation one. We want then to evaluate 
\begin{equation}
\dfrac{\langle\prod_{i=1}^{2n}\sigma(\bar{w}_{i})\epsilon(w_{1})\rangle_{\Rn}}{\langle\prod_{i=0}^{2n}\sigma(\bar{w}_{i})\rangle_{\Rn}}=\left|\dfrac{dw_{1}}{dz_{1}}\right|^{-1}\dfrac{\langle\prod_{i=1}^{2n}\sigma(v_{i})\epsilon(z_{1})\rangle_{\C}}{\langle\prod_{i=1}^{2n}\sigma(v_{i})\rangle_{\C}}.
\end{equation}
The previous expression on the complex plane can be computed using bosonization techniques, Ref.~\cite{isingchiral}, and it can be written in the general form
\begin{equation}
\langle\prod_{i=1}^{2n}\sigma(v_{i})\epsilon(z_{1})\rangle_{\C}=\sum_{\textbf{m}}\mathcal{F}_{\textbf{m}}^{(2n;N=1)}\bar{\mathcal{F}}_{\textbf{m}}^{(2n;N=1)}.\label{correlatoreising}
\end{equation}
where: 
\begin{align*}
\mathcal{F}_{\textbf{m}}^{(2n;N=1)}=2^{-\frac{n}{2}}\prod_{i=1}^{n}(v_{2i-1}-v_{2i})^{-\frac{1}{8}} & \prod_{i=1}^{2n}(v_{i}-z_{1})^{-\frac{1}{2}}(A_{2n}^{\textbf{m}})^{-\frac{1}{2}}\times\\
 & \times\left(\sum_{\textbf{t}}\left(\prod_{i=1}^{n}t_{i}^{m_{i}}\prod_{1<i,j<n}(1-x_{i,j})^{\frac{t_{i}t_{j}}{4}}\right)\Psi_{\textbf{t}}\right),
\end{align*}
and: 
\begin{eqnarray}
 & x_{i,j} & =\dfrac{(v_{2i-1}-v_{2i})(v_{2j-1}-v_{2j})}{(v_{2i-1}-v_{2j})(v_{2j-1}-v_{2i})},\\
 & A_{2n}^{\textbf{m}} & =\sum_{\textbf{t}}\left(\prod_{i=1}^{n}t_{i}^{m_{i}}\prod_{1<i,j<n}(1-x_{i,j})^{\frac{t_{i}t_{j}}{4}}\right),\\
 & \Psi_{\textbf{t}} & =-(v_{1}-v_{2})^{\frac{1}{2}}\left[\prod_{i=2}^{n}\left(\dfrac{v_{1}-v_{2i+\frac{t_{i}-1}{2}}}{v_{1}-v_{2i-1-\frac{t_{i}-1}{2}}}\right)^{\frac{1}{2}}(v_{2i-1-\frac{t_{i}-1}{2}}-z_{1})\right].
\end{eqnarray}
After taking the limit to the "in" and "out" states for the variables $v_i$, the expression $\mathcal{F}_{\textbf{m}}^{(2n;N=1)}$ is just a function of $z_{1}$, which appears just in $\Psi_{\textbf{t}}$ as
\begin{equation}
\Psi_{\textbf{t}}=\alpha_{0}+\alpha_{1}z_{1}+\dots+\alpha_{n-1}z_{1}^{n-1},
\end{equation}
and from this, \ref{correlatoreising} is a polynomial in $z_{1}$ and $z_{1}^{\ast}$ of degree $n-1$
\begin{equation}
\langle\prod_{i=1}^{2n}\sigma(v_{i})\epsilon(z_{1})\rangle_{\C}=\dfrac{\gamma_{0;0}+\gamma_{0;1}z_{1}^{\ast}+\gamma_{1;0}z_{1}^{1}+\dots+\gamma_{n-1;n-1}\left|z_{1}\right|^{2(n-1)}}{\left|\prod_{i=1}^{i=2n}(v_{i}-z_{1})\right|}.
\end{equation}
Taking then into account the transformation factor between $\Rn$ and $\C$ we obtain the value of the correlator on the Riemann surface
\begin{equation}
\dfrac{\langle\prod_{i=1}^{2n}\sigma(\bar{w}_{i})\epsilon(w_{1})\rangle_{\Rn}}{\langle\prod_{i=1}^{2n}\sigma(\bar{w}_{i})\rangle_{\Rn}}=\left|\dfrac{\pi}{nL\sin(\pi x)}\right|\dfrac{\tilde{\gamma}_{0;0}+\tilde{\gamma}_{0;1}z_{1}^{\ast}+\tilde{\gamma}_{1;0}z_{1}^{1}+\dots+\tilde{\gamma}_{n-1;n-1}\left|z_{1}\right|^{2(n-1)}}{\left|z_{1}\right|^{n-1}}.\label{isingn}
\end{equation}
The denominator of the LHS is function only of the variables $v_{i}$ and it
will become a constant after the limit to the "in" and "out" states, contributing just as a total multiplicative factor. We can approximate the previous expression around the conical singularities looking for non integrable divergences. In the point $z_1\approx 0$ it gives
\begin{equation}
\dfrac{\langle\prod_{i=1}^{2n}\sigma(\bar{w}_{i})\epsilon(w_{1})\rangle_{\Rn}}{\langle\prod_{i=1}^{2n}\sigma(\bar{w}_{i})\rangle_{\Rn}}\approx\left|\dfrac{\pi}{nL\sin(\pi x)}\right|\dfrac{\tilde{\gamma}_{0;0}}{\left|z_{1}\right|^{n-1}},\label{eq:integral1}
\end{equation}
and in $z_1\rightarrow \infty$ 
\begin{equation}
\dfrac{\langle\prod_{i=1}^{2n}\sigma(\bar{w}_{i})\epsilon(w_{1})\rangle_{\Rn}}{\langle\prod_{i=1}^{2n}\sigma(\bar{w}_{i})\rangle_{\Rn}}\approx\left|\dfrac{\pi}{nL\sin(\pi x)}\right|\tilde{\gamma}_{n-1;n-1}\left|z_{1}\right|^{n-1}.\label{eq:integral2}
\end{equation}
Integrals of \ref{eq:integral1} and \ref{eq:integral2} are both
finite in the regions around the conical singularities and no more cut-offs are needed. This term exhibits the scaling $\left(\dfrac{L\sin(\pi x)}{\pi}\right)^{-1}$, which is negligible compared to the one of the unusual corrections. This demonstrates that the corrections to the scaling of the entanglement entropy of the excited states have the same geometry dependent scalings as the ground state ones.

\section{Numerical Results}

\subsection{The XX model and its excitations}

In order to provide numerical results for our theoretical predictions,
we will perform numerical computation of the entanglement entropy
of the excited states of the XX model. This model has been extensively
studied in the literature because it can be mapped into free fermions thanks to the Jordan-Wigner transformations.\\
The Hamiltonian of the XX model reads
\begin{equation}
H_{XX}=-\frac{1}{2}\sum_{j=1}^{L}\left(\sigma_{j}^{x}\sigma_{j+1}^{x}+\sigma_{j}^{y}\sigma_{j+1}^{y}\right)+\frac{J}{2}\sum_{j}\sigma_{j}^{z},\label{eq:XX}
\end{equation}
where we used the nearest neighbors coupling as unit of energy,
$J$ is the strength of the magnetic field, $\sigma^{i}$
are the Pauli matrices and we assume periodic boundary conditions
(PBC) on the chain, $\sigma_{1}^{i}=\sigma_{L}^{i}$. This model can be mapped into a free fermionic theory thanks to the Jordan Wigner transformation
\begin{equation}
c_{l}=\left(\prod_{i=0}^{l-1}\sigma_{i}^{z}\right)\sigma_{l}^{-},
\end{equation}
where $\sigma_{i}^{\pm}$ are the ladder operators on the $i$-th site and the $c_{i}$ operators respect the standard fermionic statistic. In the Fourier space the Hamiltonian takes a diagonal form
\[
H_{XX}=\sum_{k}\left(J-2\cos\left(\frac{2\pi k}{L}\right)\right)c_{k}^{\dagger}c_{k}.
\]
 where we can identify the dispersion relation $E_{k}=J-2\cos\left(\frac{2\pi j}{L}\right)$ of the fermions.\\
The boundary condition on the fermionic problem are
\begin{equation}
c_{L+1}=(-1)^{n_{\downarrow}}c_{1},
\end{equation}
where $n_{\downarrow}$ is the number of down spins defined as
\begin{equation}
n_{\downarrow}=L-\sum_{j=0}^{L-1}\dfrac{s_{j}^{z}-1}{2},
\end{equation}
this means that the periodic conditions on the spin model and the fermionic one are not the same.\\
The model \ref{eq:XX} has different ground states depending on the values of $J$ , in the case
$J>2$ all the eigenvalues are positive definite and the ground state
is the classical state $\prod_{i}\ket{\uparrow_{i}}$, represented by the fermionic vacuum. For smaller values of the magnetic fields, $0<J<2$,
the spectrum has some eigenvalues with $E_{k}<0$ and the ground state
is composed by a Fermi Sea of all the particles with $k<k_{F}$
\begin{equation}
\prod_{m_{j}\le n_{F}}c_{m_{j}}^{\dagger}\ket{0}.\label{stati}
\end{equation}
where $k_{F}=\nu\pi=\frac{2\pi k_{c}}{L}=\arccos\left(\frac{J}{2}\right)$
is the Fermi momentum, $k_{c}=\left[\frac{L}{2\pi}\arccos\left(\frac{J}{2}\right)\right]$
is the number of states in the Fermi sea and $\nu$ is the filling factor.\\ 
Following Ref.~\cite{Peschel}, the Rényi entropy can be computed exactly and they are
\begin{equation}
S^{(n)}=\dfrac{1}{1-n}\sum_{i}^{l}\ln\left[\lambda_{i}^{n}+(1-\lambda_{i})^{n}\right],\label{eq:Rényieigen}
\end{equation}
where $\lambda_{i}$ are the eigenvalues of the correlation matrix $A_{ij}$
\[
A_{ij}=\bra{GS}c_{i}^{\dagger}c_{j}\ket{GS}
\]
In the thermodynamic limit, (\ref{eq:Rényieigen}) reproduces perfectly
the CFT predictions expected by Ref.~\cite{PasqualeCFT}
\begin{equation}
S_{gs}^{(n)}(x)=\frac{1}{6}\left(1+\frac{1}{n}\right)\ln\left[\frac{L}{\pi}\sin\left(\pi x\right)\right]+c_{n}^{\prime}
\end{equation}
The central charge is $c=1$ because the quantum field theory corresponding to the
XX chain is a free boson. Besides this continuum behavior, corrections
of the type $L^{-\frac{2}{n}}$ have been found. They correspond to
the unusual corrections to the scaling for a spinless relevant operator defined by $\Delta=1$.\\
The entanglement entropy of the excited states can be obtained in
the same way as for the ground state. Following Ref.~\cite{Berganza,BerganzaPRL}, we will study a particular type of excitations of the XX model which is the standard particle-hole excitation. It consists in the promotion of a fermionic particle from the Fermi sea to the free states above it
\begin{equation}
\ket{\text{e-h}}=c_{(n_{F}+1)/2}^{\dagger}c_{(n_{F}-1)/2}\ket{GS},
\end{equation}
in the continuum limit, this excitation corresponds to the application of $\Upsilon=\imath\partial\phi$ over the ground state represented by the Fermi sea. The entanglement entropy of this state is different from the one of the ground state Ref.~\cite{Berganza,BerganzaPRL,Pasqualedeterminante}.\\
The correlation matrix for this state is 
\begin{equation}
\tilde{A}_{ij}=\bra{e-h}c_{i}^{\dagger}c_{j}\ket{e-h}=\bra{GS}c_{(n_F+1)/2}c^\dagger_{(n_F-1)/2} c^\dagger_i c_j c_{(n_{F}+1)/2}^{\dagger}c_{(n_{F}-1)/2} \ket{GS},
\end{equation}
passing in the Fourier space we can compute this correlation function using the standard fermionic anti-commutation relations and the fact that the ground state is a Fermi sea. The correlation matrix computed for this state is
\begin{equation}
\tilde{A}_{ij} = \frac{1}{L}\left[\sum_{k\in\Omega_{\nu}}\e^{\frac{2\pi\imath}{L}(i-j)k}-\e^{\frac{\pi\imath}{L}(n_{F}-1)}+\e^{\frac{\pi\imath}{L}(n_{F}+1)}\right]
\end{equation}
where $\Omega_{\nu}$ are the possible quantum numbers for $k$. The Rényi entropy is a function of the eigenvalues of the correlation matrix $\tilde{A}_{ij}$
\[
\tilde{S}^{(n)}=\dfrac{1}{1-n}\sum_{i}^{l}\ln\left[\tilde{\lambda}_{i}^{n}+(1-\tilde{\lambda}_{i})^{n}\right].
\]
Once we have computed the Rényi entropy for the ground state and for the excited state, the function $F^{(n)}$ is simply
(\ref{eq:in}) 
\begin{equation}
F^{(n)}(x)=\e^{(1-n)\left(\tilde{S}^{(n)}-S^{(n)}\right)}.
\end{equation}

\subsection{Scaling of the corrections}

\begin{figure}[H]

\subfloat{\includegraphics[width=0.5\textwidth]{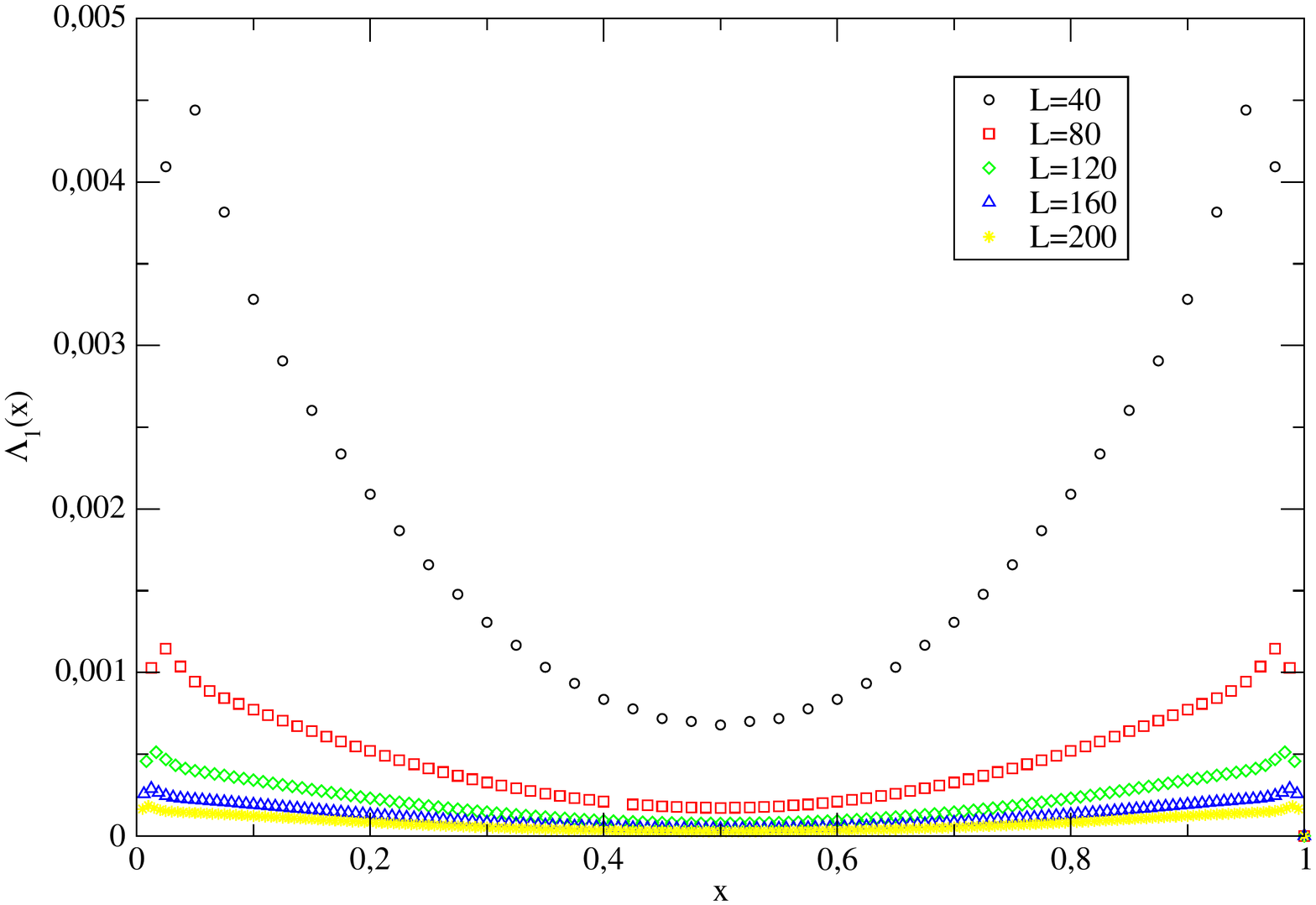}~\includegraphics[width=0.5\textwidth]{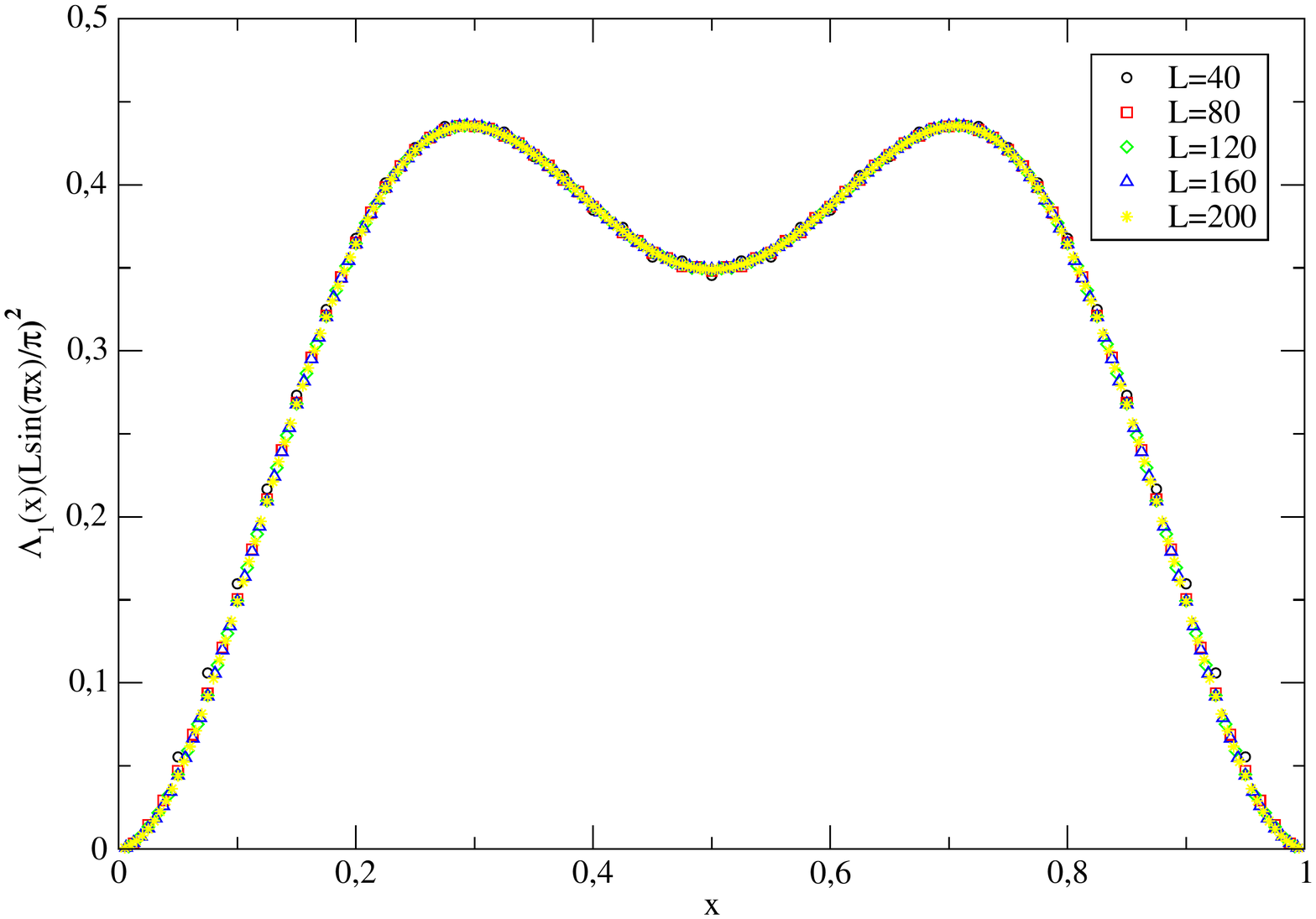}

}\\
\subfloat{\includegraphics[width=0.5\textwidth]{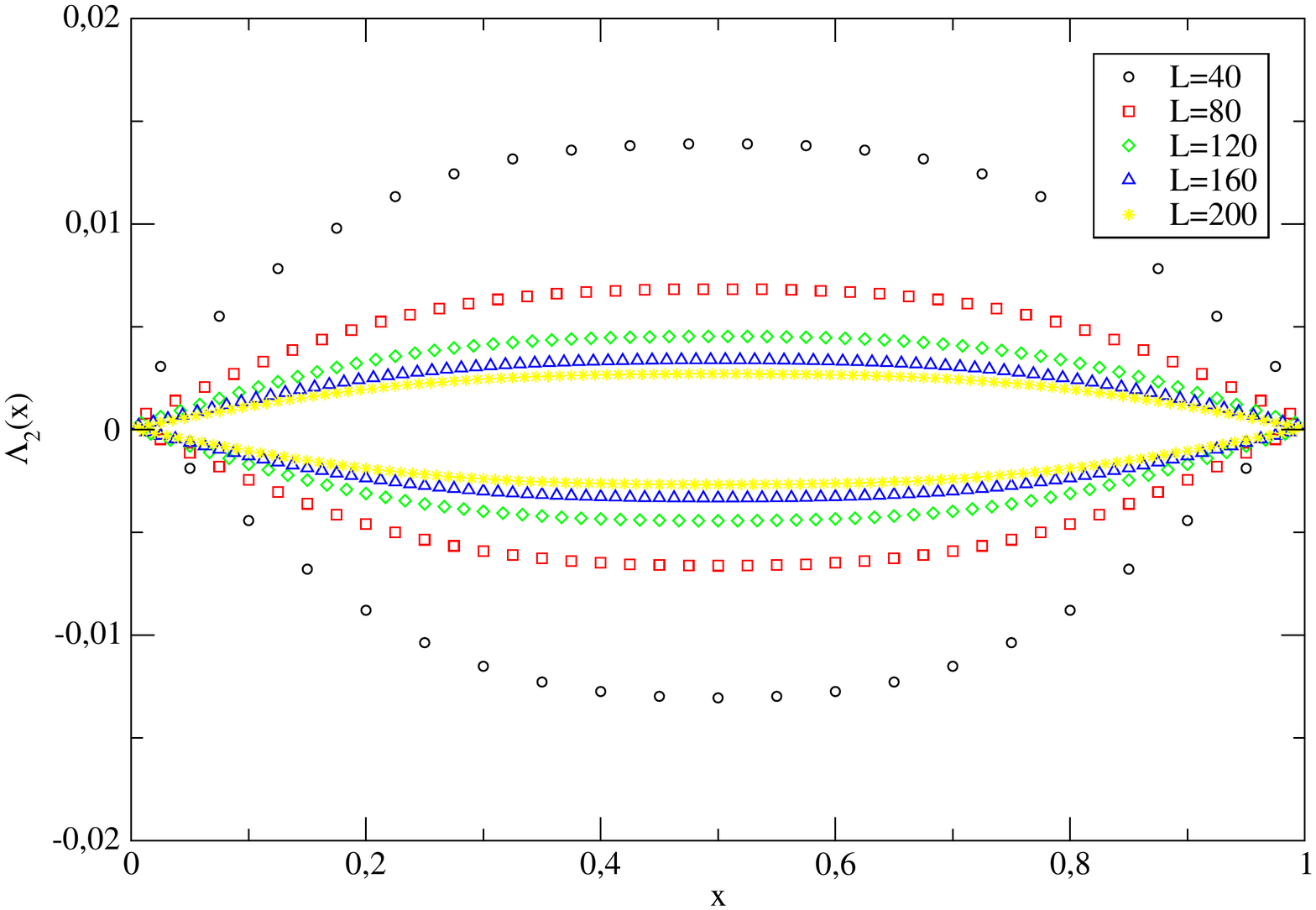}~\includegraphics[width=0.5\textwidth]{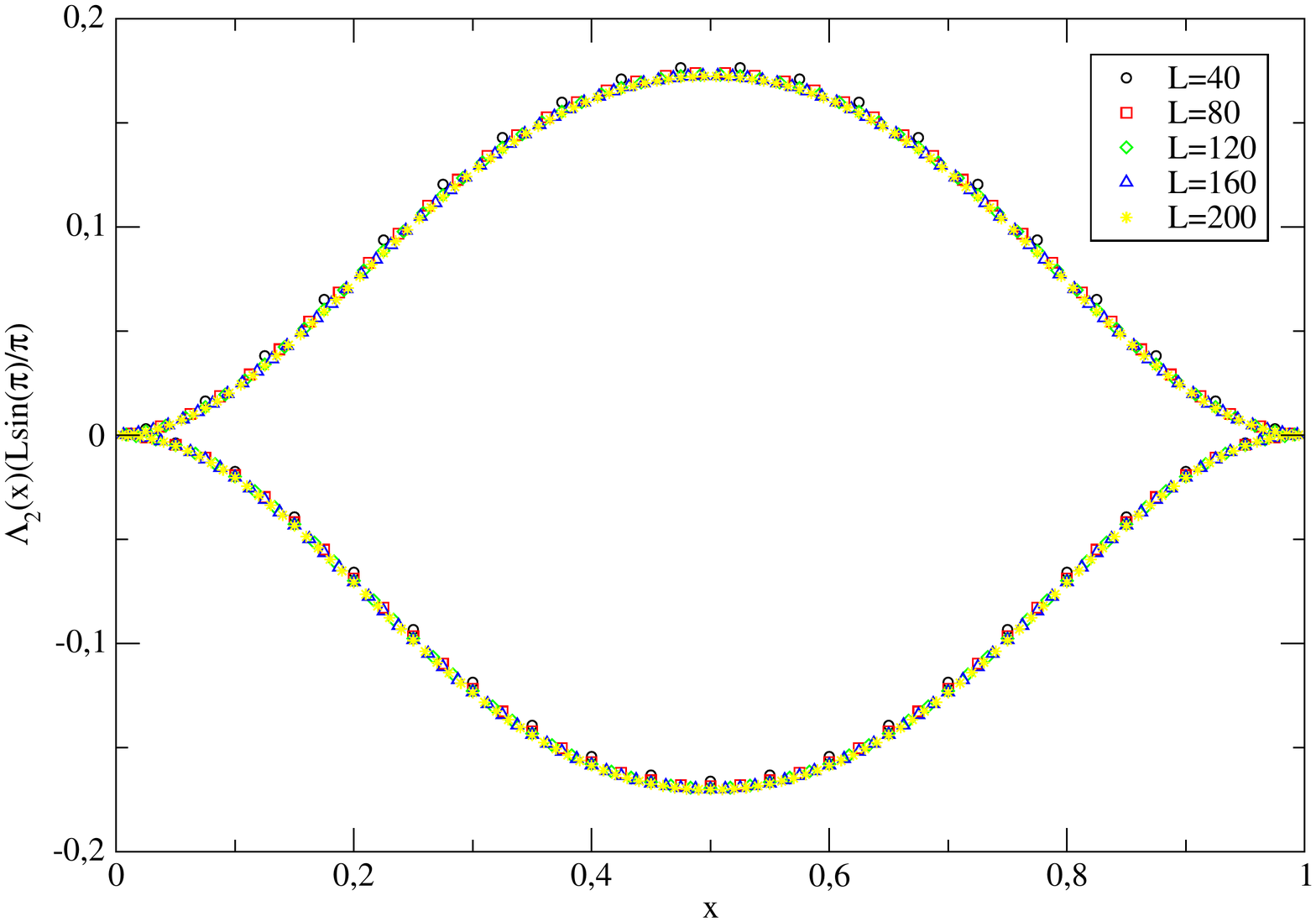}

}\\
\subfloat{

\includegraphics[width=0.5\textwidth]{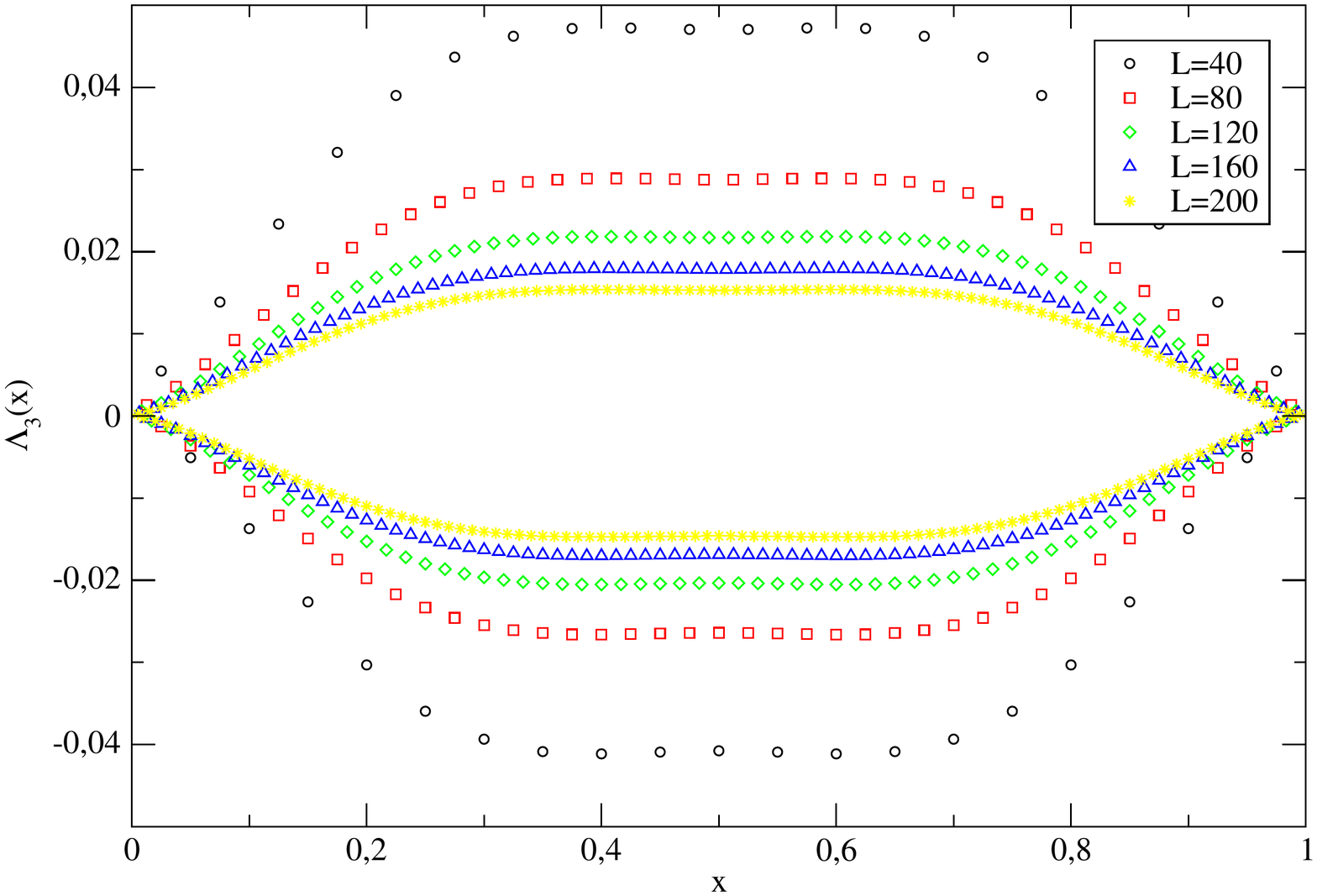}~\includegraphics[width=0.5\textwidth]{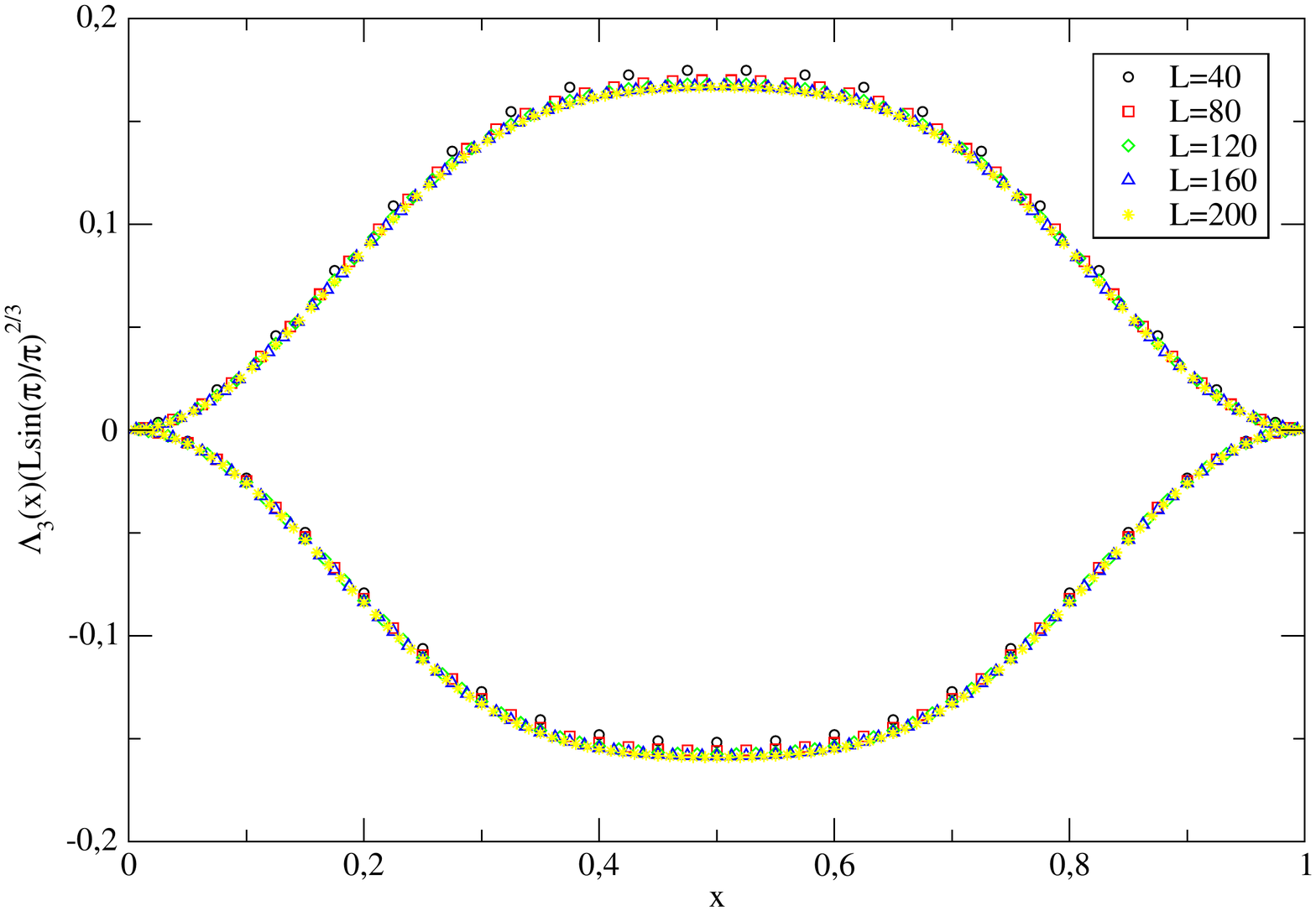}

}\caption{\footnotesize{Plots of $\Lambda_{n}$ (left) and $\left(L\sin\left(\pi x\right)/\pi\right)^{\frac{2}{n}}\Lambda_{n}$
(right) computed for the particle-hole excited state in the half-filling XX models for different values of the system size
$L$. From the top to the bottom $n=1$, $n=2$ and $n=3$.} \label{fig:half-filling}}

\end{figure}

The aim of our study is the analysis of the scaling of the corrections to the continuum CFT behavior. This is possible from the study of the dependence on the system size $L$ of the quantity
\begin{equation}
\Lambda_{n} = F^{(n)}-F_{\text{CFT}}^{(n)},
\end{equation}
where $F^{(n)}$ is obtained from the eigenvalues of the correlation matrix and
$F_{CFT}^{(n)}$ is the one obtained from analytic computations. These computations are performed using the exact expression of the correlator of $2n$ $\imath\partial\phi$ operators found in Ref.~\cite{Berganza,BerganzaPRL} and in particular its analytic continuation to real values of $n$ found in Ref.~\cite{Pasqualedeterminante,Pasqualedeterminante2}. Our theoretical analysis bases on the breaking of the conformal symmetry around the conical singularities predicts that $\Lambda_{n}$ scales as
\begin{equation}
\Lambda_{n}\sim\left(\dfrac{L\sin(\pi x)}{\pi}\right)^{-\frac{2}{n}}.
\end{equation}
In Fig.~\ref{fig:half-filling} we present the plots of $\Lambda_{n}(x)$
and the ones of $\left(L\sin(\pi x)/\pi\right)^{\frac{2}{n}}\Lambda_{n}(x)$ for
different XX systems at half-filling. Data from systems with different values of $L$ are presented in the same
plot for every value of $n$ to make an easier comparison between them. It is possible to see that the function $\Lambda_{n}(x)$
for fixed $n$ depends on the size of the system $L$ and it goes
to zero with $L\rightarrow\infty$. This represents the fact that
in the thermodynamic limit the CFT result is exact and the corrections vanish. On the other side,
$\left(L\sin(\pi x)/\pi\right)^{\frac{2}{n}}\Lambda_{n}(x)$ is no longer a function of $L$ and it depends just on the variable $x=l/L$ and on the filling, as we can see in the next figure.

%
%
%
%
%
%
%
%

\begin{figure}[H]

\subfloat{

\includegraphics[width=0.5\textwidth]{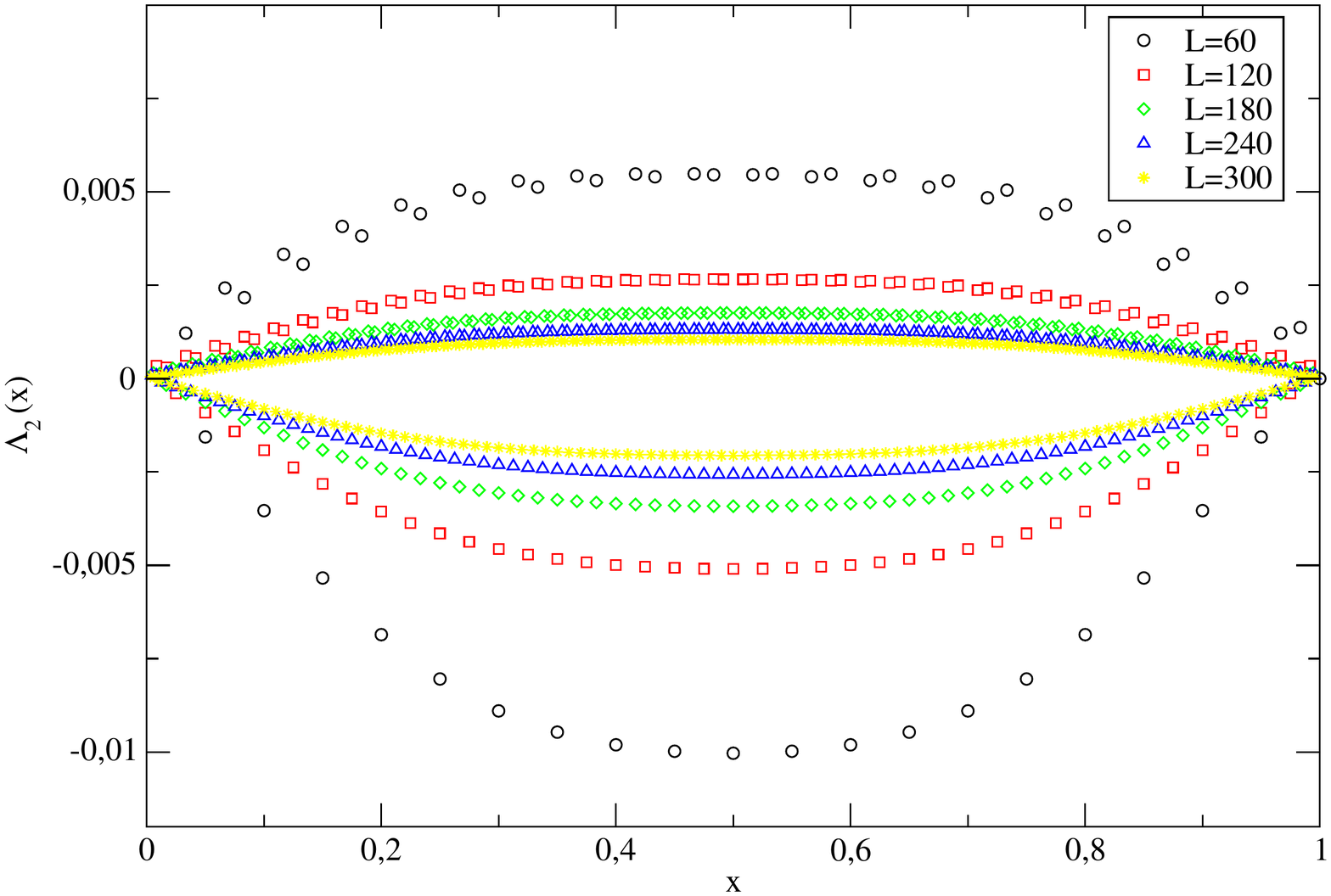}~\includegraphics[width=0.5\textwidth]{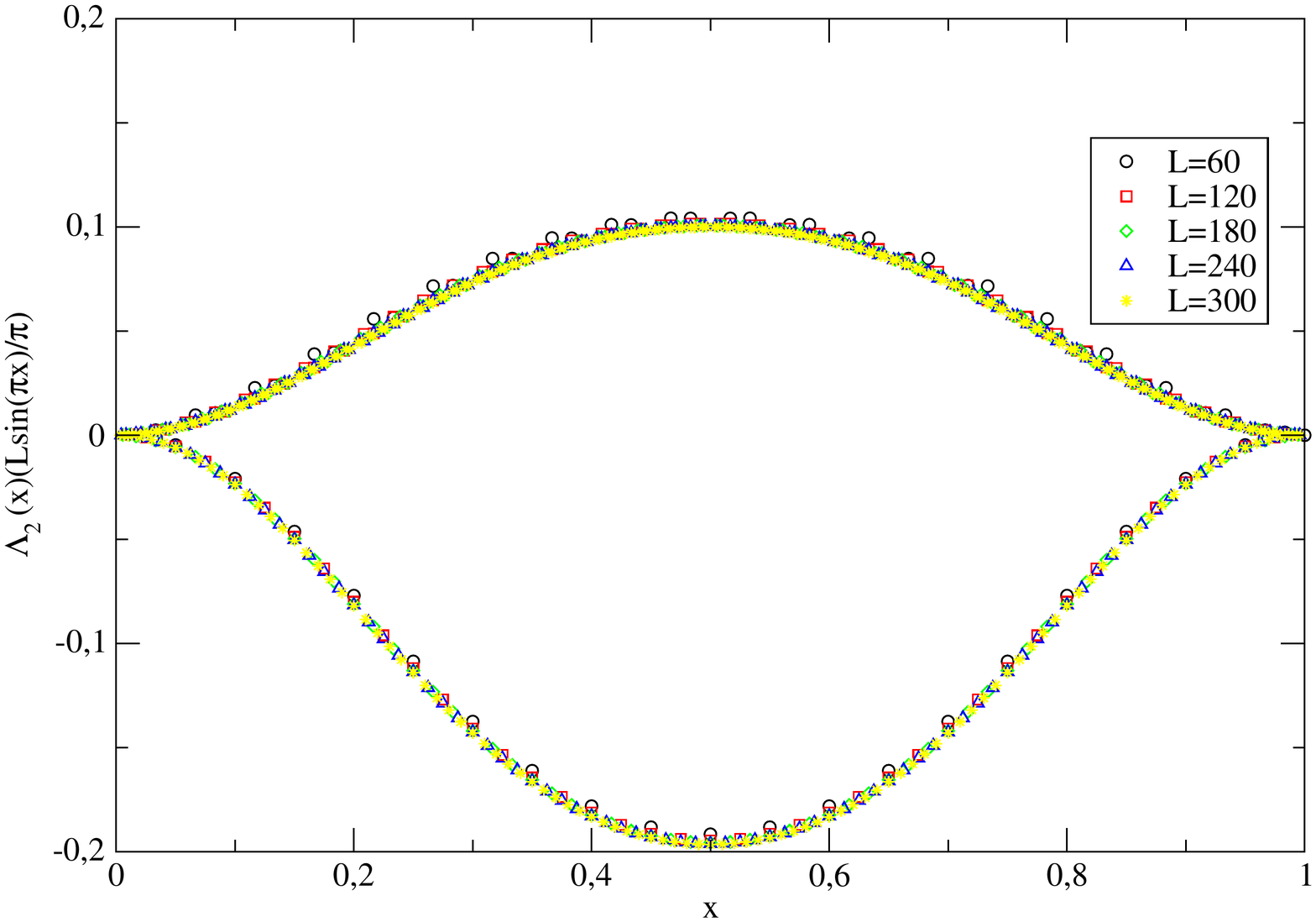} 

}\\
\subfloat{\includegraphics[width=0.5\textwidth]{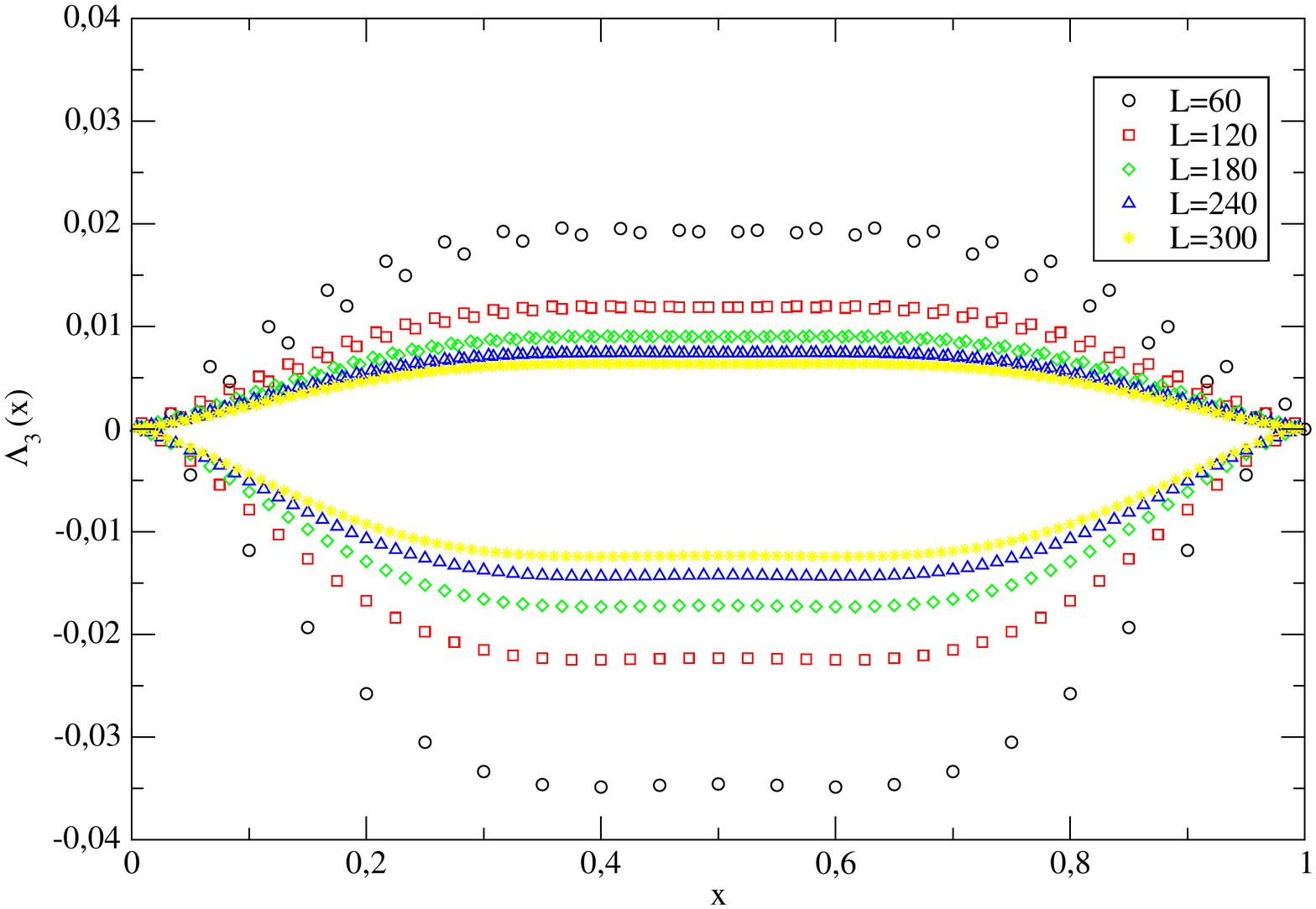}~\includegraphics[width=0.5\textwidth]{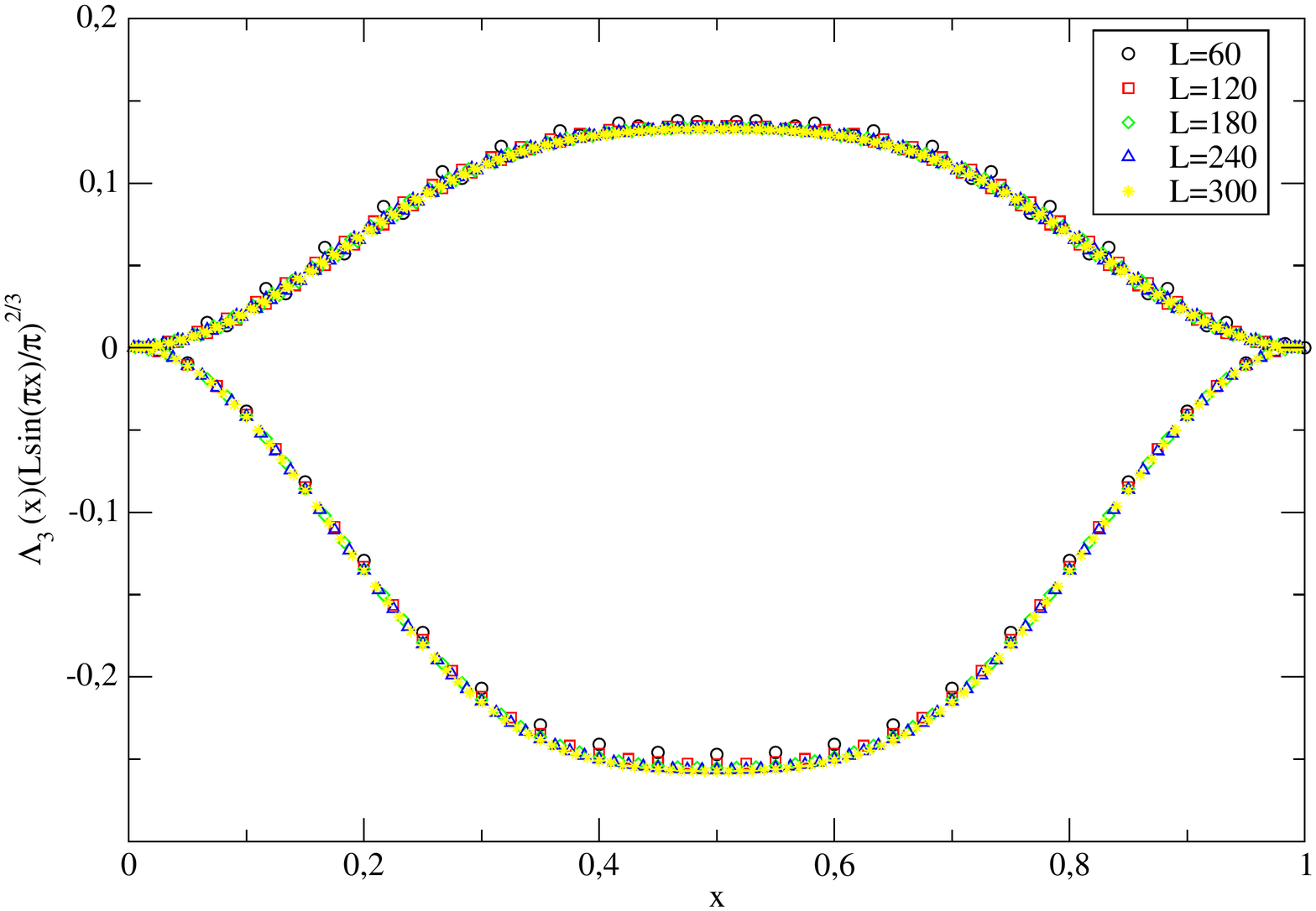} }\\
\subfloat{\includegraphics[width=0.5\textwidth]{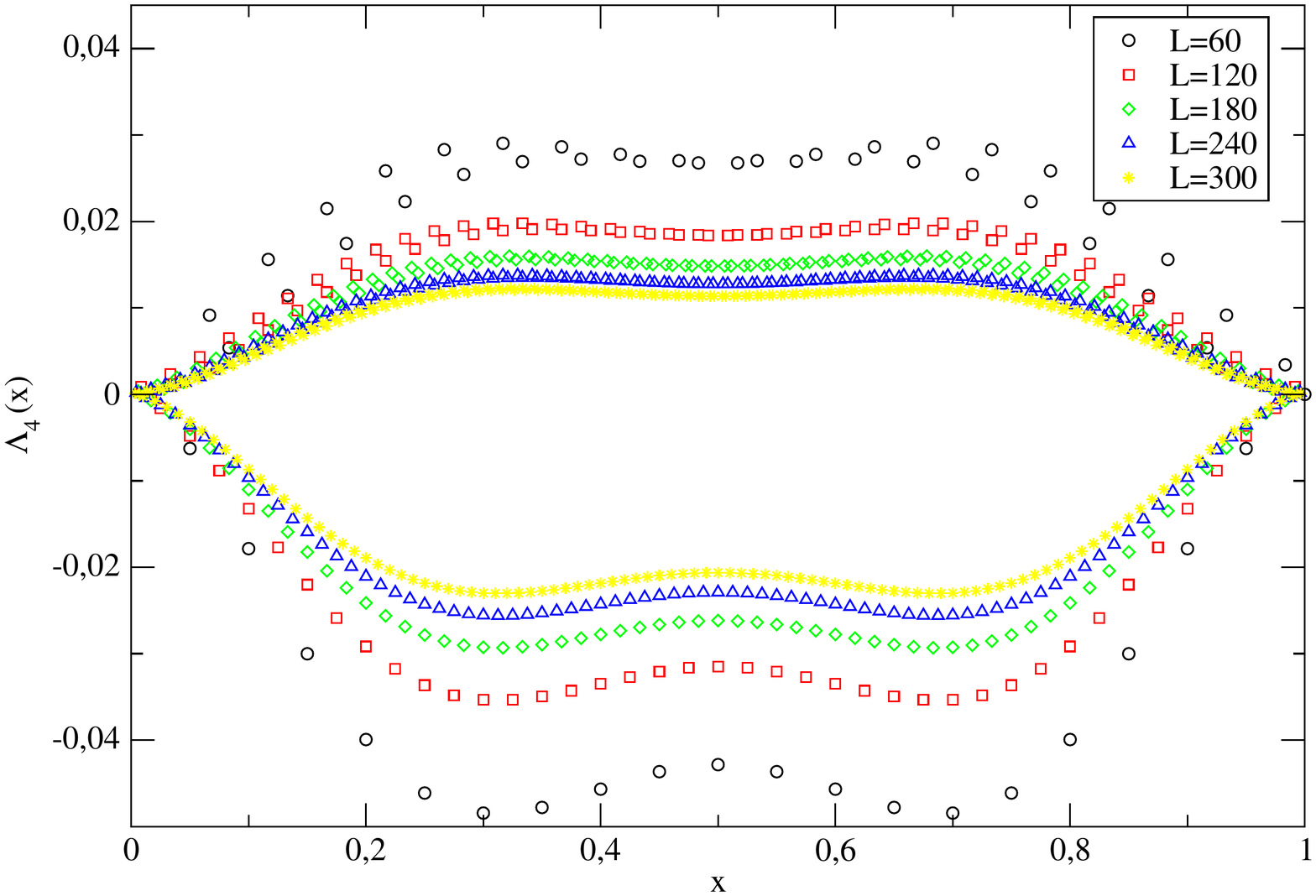}~\includegraphics[width=0.5\textwidth]{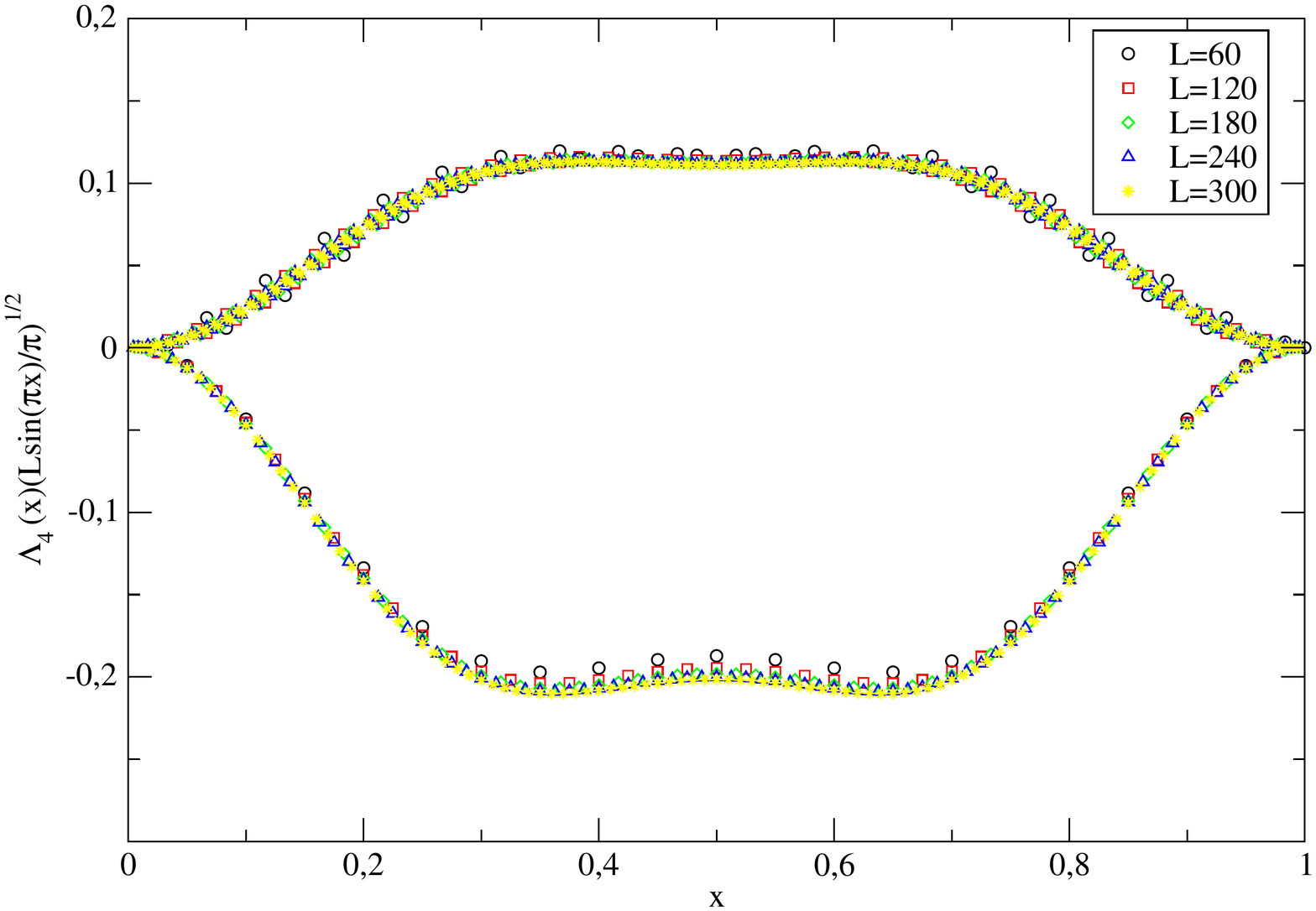} }\caption{\footnotesize{Plots of $\Lambda_{n}$ (left) and $\left(L\sin\left(\pi x\right)/\pi\right)^{-\frac{2}{n}}\Lambda_{n}$
(right)  computed for the particle-hole excited state in the $\nu=1/3$ XX models for different system size $L$.
From the top to the bottom $n=2$, $n=3$ and $n=4$.} \label{fig:filling13}}

\end{figure}

In Fig.~\ref{fig:filling13} the same quantities as before are plotted for
$\nu=1/3$ XX models for different values of the system size $L$. The scaling of $\Lambda_{n}$
is still the same as in the case $\nu=1/2$, but a comparison between
$\Lambda_{n}(L\sin\left(\pi x\right)/\pi)^{\frac{2}{n}}$ for $\nu=1/3$
and $\nu=1/2$ shows that they are not the same.

%
%
%
%
%
%
The previous analysis allows us to write the corrections in a more general way factorizing the scaling factor depending on $L$ from the function $\mathcal{F}_n(x;\nu)$ which depends just on $x=l/L$ and on the filling factor.
\[
\Lambda_{n}(x)=\left(\frac{L\sin\left(\pi x\right)}{\pi}\right)^{-\frac{2}{n}}\mathcal{F}_{n}(x;\nu)
\]
We will study the function $\mathcal {F}_n \left(x;\nu \right)$ in the next section, comparing the functions $\Lambda_n$ found for the excited states with the ones computed exactly in the ground state case.

\subsection{Comparison with non CFT results}

\begin{figure}[H]
\centering{}\includegraphics[width=0.49\textwidth]{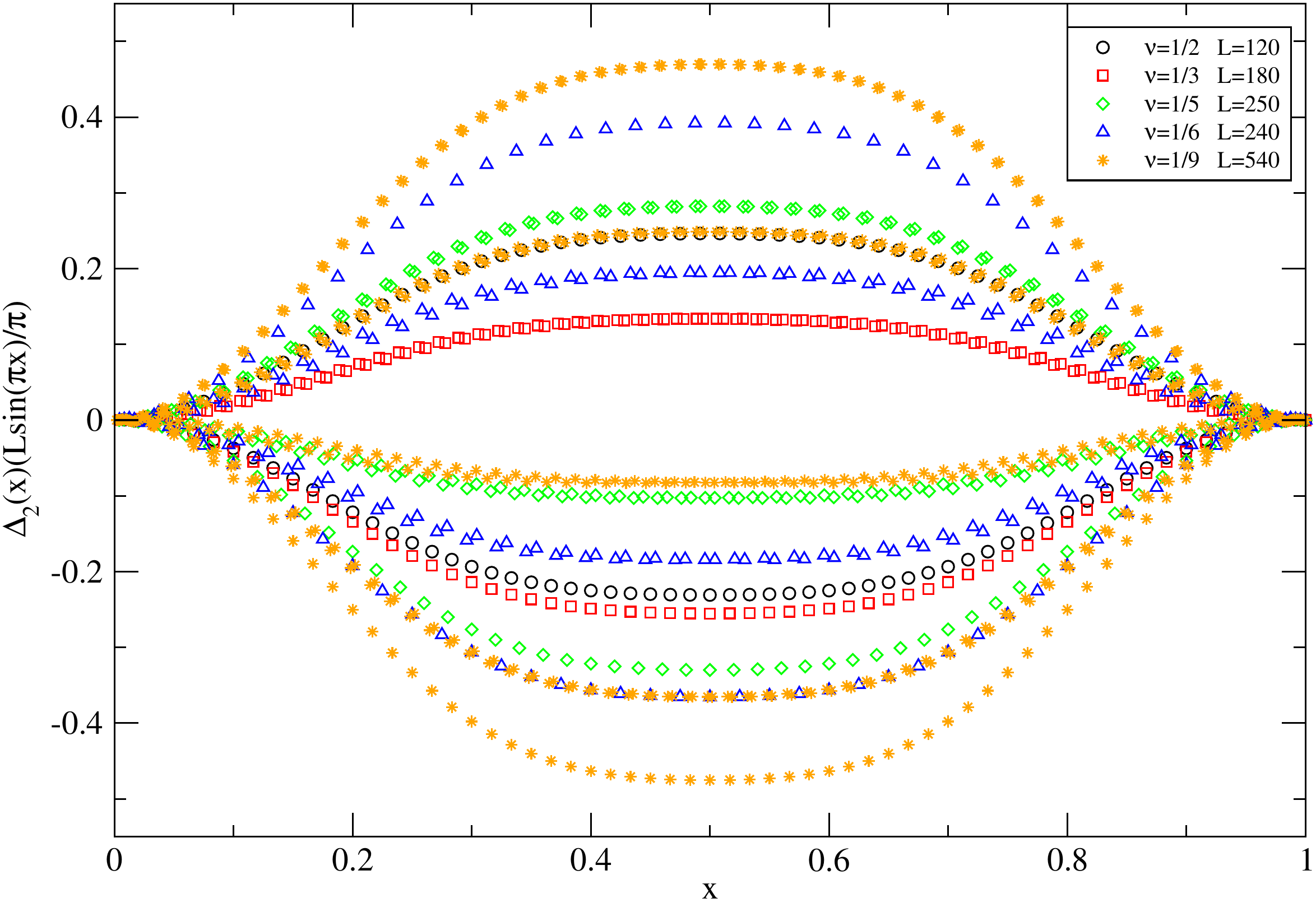}
\includegraphics[width=0.49\textwidth]{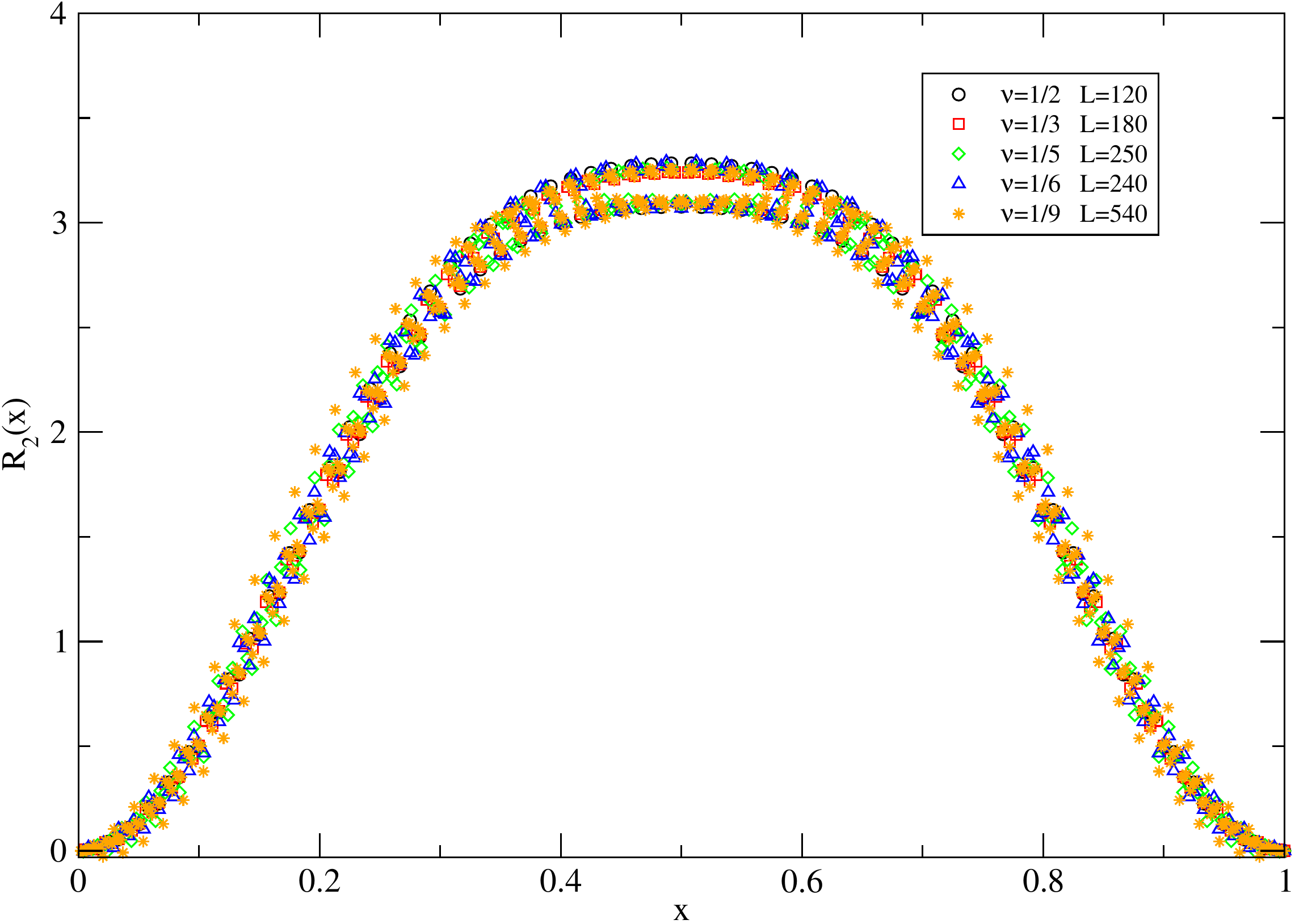} \caption{\footnotesize{Left: Plots of the function $\Lambda_{2}\left(L\sin\left(\pi x\right)/\pi\right)^{-1}$ for different values of $\nu$ and $L$. Right: Plot of $R_{2}(x)$
for different values of $\nu$ and $L$. The excited state studied is the particle-hole}\label{fig:filldiff}}
\end{figure}

The entanglement entropy of the XX model has been computed in several methods different from the 
replica trick one. In particular, in Ref.\cite{correzzioni,correzzioni2}
an analytic expression for the corrections to the continuum theory $d_{n}(N)\equiv S_{n}(N)-S_{n}^{a\rightarrow0}$
can be found. This result has been obtained with an explicit computation of the Rényi entropy which uses the properties of the Toeplitz determinants, see Ref.~\cite{351882944}.
\begin{equation}
d_{n}(N)=\dfrac{2\cos(2k_{F}l)}{1-n}\left(2N\sin(\pi l/L)\right)^{-2/n}\left[\dfrac{\Gamma\left(\frac{1}{2}+\frac{1}{2n}\right)}{\Gamma\left(\frac{1}{2}-\frac{1}{2n}\right)}\right]^{2}.\label{eq:corr}
\end{equation}
Where it is easy to recognize the scaling factor of the unusual corrections to the scaling for the ground state with $\Delta=1$. The method used in Ref.~\cite{correzzioni,correzzioni2} cannot be applied to the excited states, in this case our method is the only one that can predict the correct scaling of these corrections. We want now to study the dependence of these corrections on the filling factor $\nu$ using the quantity
\begin{equation}
R_{n}(x)=\dfrac{(1-n)(2\pi)^{2/n}}{2\cos(2k_{F}l)\left[\dfrac{\Gamma\left(\frac{1}{2}+\frac{1}{2n}\right)}{\Gamma\left(\frac{1}{2}-\frac{1}{2n}\right)}\right]^{2}}\mathcal{F}_{n}(x;\nu).\label{formulagamma}
\end{equation}
If the dependence on the filling $\nu$ is the same in both $d_{n}(x)$
and $\mathcal{F}_{n}(x;\nu)$, their ratio will no longer depend on
$\nu$ and we the function $R_{n}(x)$ will be independent of $\nu$.
In Fig.~\ref{fig:filldiff} we plot the functions $\Lambda_{2}$ and $R_{2}$ obtained
for different values of the $\nu$ and $L$ parameters. It is possible to see that
all the $R_{2}(x)$ functions collapse on the same curve. 

\begin{figure}[H]
\begin{centering}
\includegraphics[scale=0.5]{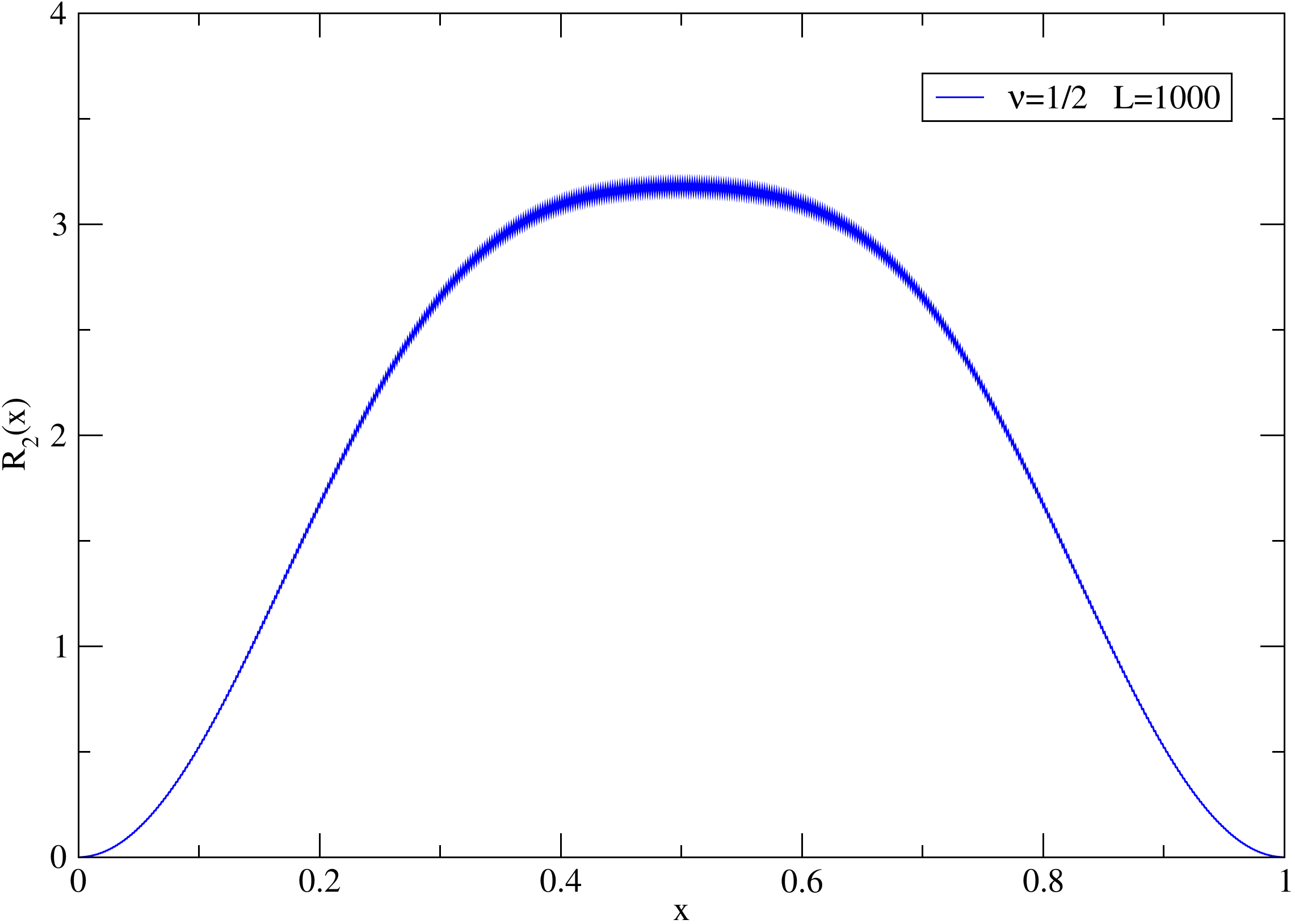} 
\par\end{centering}

\caption{\footnotesize{Plot of the function $R_{2}(x)$ obtained using a half-filled chain
of length $L=1000$. The excited state is the particle-hole state.}\label{fig:largeN}}
\end{figure}

In Fig.~\ref{fig:largeN} we plot $R_{2}(x)$ for a much bigger system
than the previous ones. This demonstrates that the residual oscillations
in Fig.~\ref{fig:filldiff} are due to finite size effects and they disappear in the thermodynamic limit.
This confirms that $R_{n}(x)$ does not depend neither on $L$ neither on $\nu$. 
We demonstrated that the corrections to the entanglement entropy of the excited states are the same as the one of the ground state up to a function of $l/L$, we demonstrated that it is possible to write these corrections in the general form
\begin{equation}
\Lambda_{n}=d_n\left( N \right)R_n(x),
\end{equation}
where $d_n(N)$ are the corrections to the continuum theory in the ground state case and the function $R_n(x)$ is an unknown model dependent function which depends just on $l/L$.

\section{Conclusions \& Outlooks}

We demonstrated that the unusual corrections to the scaling of the entanglement
entropy in the excited states of the conformal field theory are due to the
local breaking of the conformal invariance around the conical singularities
of the Riemann surface, as in the ground state case. Many terms contribute
to these corrections but the dominant one scales as $\left(L\sin(\pi x)/\pi\right)^{-\frac{2\Lambda}{n}}$
where it is possible to admire the geometrical dependence of the scaling of these corrections. We checked
our predictions using exact numerical computations of the entanglement entropy
of the excited states of the XX model finding a perfect agreement. By comparing our numerical data with the exact analytical expression for these corrections in the ground-state found in Ref.~\cite{correzzioni}, we find that the dependence on the filling factor is the same both in the ground and the excited states.\\
In the future it would be interesting to compute the function $R_{n}(x)$ to see if it is
model dependent or not. This can be done in the random XX model, where the entanglement entropy can be studied analytically for the ground and the excited states, see Ref.~\cite{hopping1,hopping2}. Moreover, the study of the corrections to the CFT results can be done for Rieman surfaces different from the one considered in this work. The entanglement negativity between two disjoints subsystem $A_1$ and $A_2$ of a system $S$ has in fact proven to be described by a richer structure than the single interval case, Ref.~\cite{neg1,neg2,neg3}, and the CFT result could be affected again by more complicated geometry dependent corrections.\\
\newline
\textit{Acknowledgments} \footnotesize{We thanks P. Calabrese, E. Ercolessi, L. Taddia, A. Fabbri, P. Mattioli for very helpful discussions. The author is supported by the ERC (FP7/2007-2013 Grant Agreement No. 256294).}
 

\end{document}